\documentclass[12pt]{article}

\pdfoutput=1

\usepackage[top=80pt,bottom=85pt,left=85pt,right=85pt]{geometry}
\usepackage{amsmath,amssymb,graphicx,float,color,tikz,cite,savesym,wasysym}
\usepackage{caption}
\usepackage{subcaption}
\usepackage[debug,pageanchor=false]{hyperref}
\definecolor{link}{rgb}{.8,.15,.1}
\hypersetup{colorlinks=true,linkcolor=link,citecolor=link,urlcolor=link,linktocpage}

\newcommand{\del}{\partial}

\makeatletter
\@addtoreset{equation}{section}
\makeatother

\savesymbol{iint}
\savesymbol{iiint}
\restoresymbol{TXF}{iint}
\restoresymbol{TXF}{iiint}
\newcommand{\tmem}[1]{{\em #1\/}}

\begin{document}

	\begin{titlepage}

	\begin{center}

	\vskip .5in 
	\noindent

	{\Large \bf{New de Sitter Solutions in Ten Dimensions\\ and Orientifold Singularities}}

	\bigskip\medskip

	 Clay C\'{o}rdova,$^1$ G.~Bruno De Luca,$^{2}$ Alessandro Tomasiello$^{2}$\\

	\bigskip\medskip
	{\small 
	\vspace{.3cm}
$^1$ Kadanoff Center for Theoretical Physics  \& Enrico Fermi Institute\\
University of Chicago, Chicago, IL 60637, USA 
\\	
	\vspace{.3cm}
$^2$ Dipartimento di Fisica, Universit\`a di Milano--Bicocca, \\ Piazza della Scienza 3, I-20126 Milano, Italy \\ and \\ INFN, sezione di Milano--Bicocca
	
		}

	\vskip .5cm 
	{\small \tt clayc@uchicago.edu, g.deluca8@campus.unimib.it,  alessandro.tomasiello@unimib.it}
	\vskip .9cm 
	     	{\bf Abstract }
	\vskip .1in
	\end{center}

	\noindent In previous work, we found ten-dimensional solutions to the supergravity equations of motion with a dS$_4$ factor and O8-planes. We generalize this analysis and obtain other solutions in the same spirit, with an O8$_+$ and an O6$_-$.  We examine our original solutions in more detail, focusing in particular on the O8$_-$ singularities and on the issues created by their boundary conditions. We also point out some previously known supersymmetric AdS solutions with the same local behavior at their O8$_-$ singularity.

	\noindent

	\vfill
	\eject

	\end{titlepage}

\tableofcontents
	
\section{Introduction} 
\label{sec:intro}

It is a long-standing challenge in string theory to construct solutions with a positive cosmological constant.  An essential complication is that within the low-energy supergravity limit there are no-go arguments \cite{gibbons-nogo, dewit-smit-haridass, maldacena-nunez} that forbid de Sitter compactifications using only ingredients obeying standard energy conditions.  

Because of this, any putative de Sitter solution must in some way violate the classical supergravity approximation.  For instance, one may make use of corrections to the two-derivative supergravity equations.  Alternatively one can try to construct solutions using semiclassical objects, orientifolds, that have negative tension and violate the assumed energy conditions.  The latter class of constructions also takes us beyond the supergravity approximation.  Close to the orientifolds the curvature and dilaton often become large and stringy corrections again become important.  

Several classes of de Sitter models have been proposed using these ideas; see for example \cite{kklt,balasubramanian-berglund-conlon-quevedo, silverstein-simple, caviezel-koerber-kors-lust-wrase-zagermann, danielsson-koerber-vanriet}.  However, because of the importance of stringy or quantum corrections, these solutions are typically not under parametric control; moreover the O-planes are often smeared (although see  \cite{dong-horn-silverstein-torroba,blaback-danielsson-junghans-vanriet-wrase-zagermann} for investigations on how to unsmear them).  For these reasons, doubts have remained both over particular proposals \cite{bena-grana-halmagyi, banks-10500, Sethi:2017phn} and even over the existence any de Sitter compactifications in string theory; see  \cite{danielsson-vanriet,obied-ooguri-spodyneiko-vafa,andriot-swamp,denef-hebecker-wrase,Garg:2018reu, ooguri-palti-shiu-vafa, andriot-open,kachru-trivedi,das-haque-underwood} for a sample of recent discussions.  

Motivated by these considerations, in \cite{cordova-deluca-t-ds4} we constructed explicit de Sitter compactifications of ten-dimensional supergravity by directly solving the IIA supergravity equations of motion. Spacetime is a warped product of dS$_4$ with an internal $M_6$. The no-go arguments are evaded because of the presence of two singularities, which we identified from the behavior of the fields as being those of an O8$_+$ and an O8$_-$-plane. Of course near these singularities supergravity breaks down: in particular, both the curvature and the string coupling become large at a finite distance from the O8$_-$. To assess the validity of these solutions in string theory, one should ideally use the full string theory action, or switch to a dual description. Unfortunately neither of these options is available, and for this reason we emphasized in \cite{cordova-deluca-t-ds4} that the ultimate fate of those solutions depends on string theory corrections.  Here we do not resolve this issue, though in section \ref{sec:conclusions} below we discuss some ideas to attack these problems indirectly.

\begin{figure}[ht]
	\centering
		\includegraphics[width=6cm]{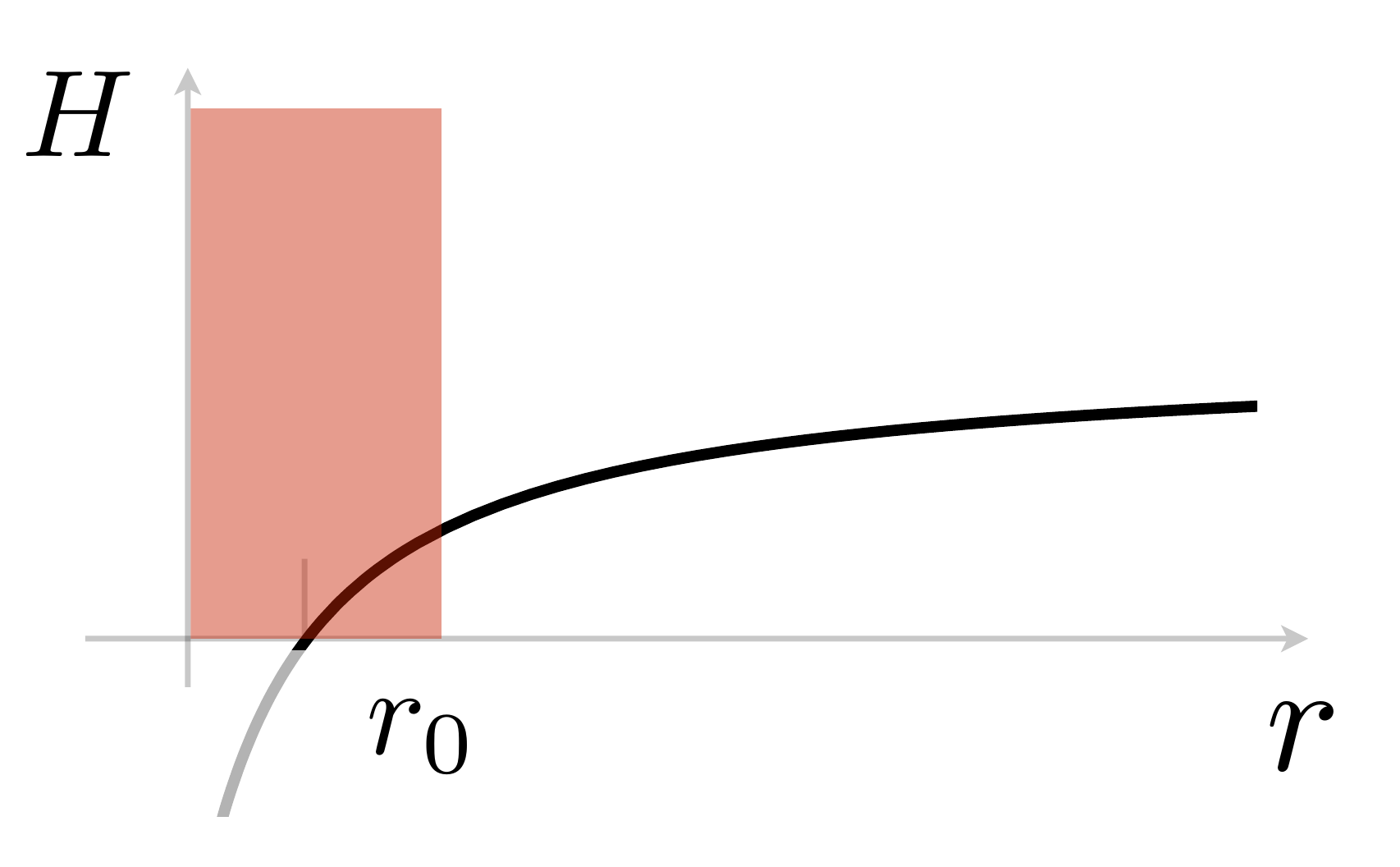}
	\caption{\small The function $H$ for O$p$-planes with $p<7$ becomes negative at some value $r=r_0$ of the radial coordinate. For $3<p<7$, curvature and string coupling become strong already for larger values of $r$, in a region schematically shown in red here.}
	\label{fig:hole}
\end{figure}

Instead, one of our main results in this paper is to construct a new class of ten-dimensional de Sitter compactifications of massive IIA.  In these new solutions, described in section \ref{sec:o8o6}, the O8$_-$ is replaced by an O6$_-$. As we review in section \ref{sec:op}, for such an object, as for any O$p_-$-plane with $p<8$, it is well-known that the supergravity solution breaks down in a ``hole'' region around the source; see Figure \ref{fig:hole}.  In absence of Romans mass, it is known how to resolve this singular behavior in M-theory, where it is replaced by the smooth Atiyah--Hitchin metric \cite{seiberg-witten-3d,hanany-pioline}; with $F_0\neq 0$ however this is not possible.  

By modifying an analytic class of AdS solutions \cite{passias-rota-t}, we will be able to find numerical de Sitter solutions where the metric takes the form
\begin{equation}\label{eq:intro}
  d   s^2_{10} = e^{2 W} d   s^2_4 + e^{- 2 W} (e^{2 \lambda_3}
  d   s^2_{\kappa_3} + d   z^2 + e^{2 \lambda_2} d
    s^2_{S^2})~,
\end{equation}
where the warp factor $W$ as well as the dilaton and functions $\lambda_{i}$ depend only on the single coordinate $z$, and $d   s^2_{\kappa_3}$ is the metric on an Einstein three-manifold with negative curvature.\footnote{Relative to our later discussion we have suppressed the gauge redundant function $Q$.}  The $z$ direction parameterizes an interval subject to an orientifold involution.  At one side there is an O8$_+$-plane, on the other side the solution terminates at the boundary of a hole, behaving locally like the O6$_-$ in flat space. There are several AdS solutions where the presence of an O6$_-$ has been argued using this hole behavior, including some with known holographic dual \cite{afrt,apruzzi-fazzi}.

In section \ref{sec:o8o8}, we take a step back from these explicit discussions and analyze in more detail the original solutions  of \cite{cordova-deluca-t-ds4} with only O8s.  Our discussion here is motivated in part by a complaint \cite{cribiori-junghans} about the O-plane singularities in \cite{cordova-deluca-t-ds4}. By using a certain combination of the supergravity equations of motion, \cite{cribiori-junghans} claimed that no dS solutions with only O8-plane singularities could exist; they resolved the apparent contradiction with \cite{cordova-deluca-t-ds4} by finding fault with the subleading behavior of the fields as a function of distance from the O8$_-$.

The argument in \cite{cribiori-junghans} assumes the validity of the supergravity equations of motion everywhere, even near the orientifolds where they obviously are invalid. Their apparent aim is to ascertain whether the solution can be trusted in supergravity, before one goes on to consider stringy corrections.  Of course as discussed above all de Sitter solutions necessarily involve some correction to the low-energy supergravity approximation.  However, such a breakdown does not settle the essential physical question of whether \emph{string theory} admits de Sitter vacua.  

The nature of the solutions of \cite{cordova-deluca-t-ds4}, as with all solutions involving orientifolds, is that the supergravity is a valid approximation in one region, while completely breaking down in another.  Of course this means that the solutions cannot be completely verified without taking into account the strong-coupling region. However, this issue cannot possibly adjudicated one way or another by using the equations of motion of supergravity.   

Nevertheless, in section \ref{sec:o8o8} we discuss the \emph{formal} problem of the behavior of the supergravity fields and equations in the strong-coupling region of an O8.  As one might anticipate, the confusion is one of boundary conditions: \cite{cordova-deluca-t-ds4} and \cite{cribiori-junghans} use two slightly different versions, imposing that a certain function have a single or double zero. This comes in turn from two different assumptions on the allowed field fluctuations near O-planes.   These issues are in fact independent of the sign of the cosmological constant.  There are purely local questions about the correct local description of O8$_-$ singularities.  In particular in appendix \ref{bdsusy} we show that certain previously constructed analytic supersymmetric AdS solutions \cite{passias-prins-t} have O8$_-$ singularities with the same boundary behavior as \cite{cordova-deluca-t-ds4}.

A priori, without input from a more fundamental theory one cannot determine which, if any of these boundary behaviors is correct.  Thus, the status of these O8s is somewhat similar to black brane solutions of supergravity where one requires input from string theory to decide which solutions and singularities are physical.  The O6 solutions of section \ref{sec:o8o6} have a similar status.  Our analysis of these ensures that they have the correct charge and leading approximation to an orientifold near the boundary of the hole region but leaves the question of whether additional boundary conditions should be imposed presently unanswered.

 
\section{Orientifolds in Supergravity} 
\label{sec:op}

Since we will have several O$p$-singularities in what follows, we start with a review of their effect in supergravity. We only need in fact O$8$s and O$6$s in this paper, but consider other values below for completeness.

The most common type of O$p$-plane is the so-called O$p_-$, which has negative tension and charge. We will mostly focus on this case, commenting only occasionally on O$p_+$-planes, which have positive tension and charge, and hence a behavior more similar to a stack of D$p$-branes.\footnote{A stack of D$p$-branes coincident with an O$p_{-}$ has an SO worldvolume gauge group, while D$p$-branes coincident with O$p_{+}$ give SP gauge groups. }  In the context of dS solutions discussed in this paper, we will always need at least one O$p_{-}$ to violate the no-go arguments of \cite{maldacena-nunez}.  We will work in string frame unless otherwise noted.

For general $p$, the O$p_-$ solution can be obtained by a modification of the D$p$ solution and reads:
\begin{equation}\label{eq:Op}
  d   s^2 = H^{- 1 / 2} d   s^2_{p + 1} + H^{1 / 2} (d  
  r^2 + r^2 d   s^2_{S^{8 - p}}) \, ,\qquad F_{8 - p} = - \frac{4 \pi^{7-p}}{v_{8-p}} \text{vol}_{S^{8 - p}}\, ,\qquad e^{\phi} = g_s
  H^{\frac{3 - p}{4}}\,.
\end{equation}
Here $d   s^2_{p + 1}$ is the space parallel to the O$p$-plane, $v_d= 2\frac{\pi^{d/2}}{\Gamma(d/2)}$ is the volume of the unit-radius $S^{8-p}$, and $H$ is a harmonic function of the transverse coordinates. 

\subparagraph{$p=8$.} 

In this case there is a single transverse coordinate, which we call $x^9$; the metric in (\ref{eq:Op}) now reads $ds^2= H^{-1/2}ds^2_8 + H^{1/2} (dx^9)^2$, with 
\begin{equation}\label{eq:HO8-flat}
	H_{\mathrm{O8}_-}= a + \frac{g_s}{2\pi} x^9\,.
\end{equation}
For $a>0$ the curvature and string coupling are finite, but at the origin $x^9=0$ the string coupling $e^\phi=g_s a^{-5/4}$ and $R\propto g_s^2 a^{-5/2}$, which may be large if $a$ is small. In particular for $a=0$ they diverge: $e^\phi\sim (x^9)^{-5/4}$, $R\sim (x^9)^{-5/2}$.\footnote{Supergravity solutions with $a=0$ are ubiquitous.  For instance they occur in well-studied supersymmetric AdS solutions (see e.g.~\cite{brandhuber-oz,bah-passias-t,passias-prins-t,dibitetto-lomonaco-passias-petri-t,dibitetto-passias}), as well as the non-supersymmetric AdS$_{8}$ solutions of \cite{cordova-deluca-t-ads8} and the dS solutions of \cite{cordova-deluca-t-ds4}. We discuss some subtleties of these O8s in section \ref{sub:recap} below. } At large $x^9$ the metric does not reduce to flat space.

An O8$_+$ is obtained by reversing the sign of the second term in (\ref{eq:HO8-flat}), so $H_{\mathrm{O8}_+}= a - \frac{g_s}{2\pi} x^9$. Because of the square roots of $H$ in (\ref{eq:Op}), the metric now loses meaning beyond a critical distance, $x^9\ge 2\pi a/g_s$, just like for a stack of D8-branes. This will not be important for our setting, in which the transverse directions are compact. For an O8$_+$ it is impossible to take $a=0$: it would take the critical distance to zero and make the solution disappear altogether.

\subparagraph{$p=7$.}

We briefly include this case for completeness, but we will not consider it in this paper. Here 
\begin{equation}\label{eq:HO7-}
	H_{\mathrm{O7}_-}= \frac{2 g_s}\pi \log\left(\frac r{r_0}\right)\,. 
\end{equation}
 As was the case for the O$8_-$, at $r\to \infty$ the metric does not reduce to the flat space metric. 
For $r<r_0$, $H$ becomes negative, and once again the metric becomes imaginary. This unphysical ``hole'' is resolved in F-theory, where the hole is revealed to contain two mutually non-local $(p,q)$-sevenbranes \cite{sen-f1}. The hole also occurs for $p<7$, where a similar resolution is not always known, and we will discuss it at greater length. 

For an O7$_+$, the sign in (\ref{eq:HO7-}) is reversed, and the metric again has a critical distance, as for the O8$_+$.

\subparagraph{$3<p<7$.}

The harmonic function is now a power law:
\begin{equation}\label{eq:HOp}
 H \equiv 1 - \left( \frac{r_0}r\right)^{7-p} \, ,\qquad (p<7)\,.
\end{equation}
The metric becomes imaginary in the ``hole''
\begin{equation}\label{eq:hole}
  r < r_0 \propto g_s\,;
\end{equation}
This hole region is a finite distance in the metric and both the curvature and string coupling diverge as one goes towards the ``boundary of the hole'' $r=r_0$: namely
\begin{equation}
	R\sim (r-r_0)^{-5/2} \, ,\qquad e^\phi\sim (r-r_0)^{\frac14(3-p)}\,.
\end{equation} 
So the solution cannot be trusted already some distance outside the hole. We will return on this issue in section \ref{sec:o8o6}. 

For O$p_+$ in the range $3<p<7$, the sign of the second term in (\ref{eq:HOp}) is reversed. Now nothing special happens at $r=r_0$; as $r\to 0$ the curvature diverges like $R\sim r^{\frac12 (3-p)}$, while the string coupling goes to zero like  $e^\phi\sim r^{\frac14\frac{p-3}{7-p}}$.

\subparagraph{$p=3$.}

The harmonic function for an O3$_-$ is still (\ref{eq:HOp}), but now the string coupling is a constant. The curvature $R$ in fact also vanishes, but the invariant $R_{MN}R^{MN}$ diverges as $\sim (r-r_0)^{-5}$ as $r\to r_0$. So $r=r_0$ is still a true singularity. 

For the O3$_+$, the sign in (\ref{eq:HOp}) is reversed; again $e^\phi$ is a constant and $R=0$. Other curvature invariants also remain finite in this case, and it was in fact argued in \cite{gibbons-horowitz-townsend} that the locus at $r\to 0$ is not a singularity, and that one can analytically continue beyond it. 

\subparagraph{$p<3$.}

This case plays no role in four-dimensional compactifications, but let's review it just the same. At $r=r_0$ the curvature again diverges as $(r-r_0)^{-5/2}$, while the string coupling goes to zero as $e^\phi\sim (r-r_0)^{\frac14(3-p)}$. For O$p_+$'s in this range, the curvature goes to zero for $r=0$ as $R\sim r^{\frac12(3-p)}$, while the string coupling diverges as $e^\phi\sim r^{\frac14\frac{p-3}{7-p}}$.


\section{O8$_+$--O6$_-$ solutions} 
\label{sec:o8o6}

In this section we construct a new type of ten-dimensional de Sitter solutions of supergravity, involving O8$_{+}$ and O6$_-$ orientifolds.  We first present a short review of the ansatz of \cite{cordova-deluca-t-ds4} and then subsequently generalize.  In section \ref{sub:ads7} we also show that our ansatz encompasses some analytically known AdS$_{4}$ solutions.

\subsection{Review of the O8$_+$--O8$_{-}$ Ansatz}

The simplest class of solutions in \cite{cordova-deluca-t-ds4} are ten-dimensional geometries with a dS$_4$ factor in spacetime, and two O8-planes. The internal geometry is $S^1 \times M_5$ and only $F_0$ flux. The $S^1$ is quotiented by an orientifold, so that one can consider an interval $I$ instead. 

The metric ansatz was 
\begin{equation}\label{eq:metriceasy}
	ds^2_{10}= e^{2W} ds^2_{\mathrm{dS}_4}+  e^{-2W}\left(dz^2+e^{2\lambda} ds^2_{M_5}\right)\,.
\end{equation}
$M_5$ is a compact Einstein manifold with negative curvature; $z$ is a coordinate on the above-mentioned $S^1/\mathbb{Z}_2\equiv [0,z_0]$. The metric coefficients $W$, $\lambda$, as well as the dilaton $\phi$ are all functions of the coordinate $z$ only. 

At the loci $z=0$ and $z=z_0$, the metric is singular. The $z=0$ singularity is very mild: the metric is continuous but non-differentiable there. We interpreted this as the backreaction of an O8$_+$-plane, which has positive tension; in fact we started our analysis by imposing boundary conditions at $z=0$ that correspond to the backreaction of an O8$_+$. These boundary conditions are not controversial in any way. We then started a numerical evolution, and at $z=z_0$ found another singularity. Closer inspection revealed that this matched the behavior of an O8$_-$ at leading order in $|z-z_0|$. This led to the claim that we had found a solution with an O8$_+$ and an O8$_-$.

Below we describe a new type of solutions, where the O8$_-$ is replaced by an O6$_-$.   We return to a critical analysis of the O8$_-$ solutions and their orientifold boundary conditions in section \ref{sec:o8o8}. In particular there we also discuss the arguments of \cite{cribiori-junghans}.

\subsection{Setup} 
\label{sub:O6-setup}

Let us first describe the allowed singularities.  We will recognize the O6$_-$ by the behavior at the boundary of its ``hole'' $r=r_0$, discussed in section \ref{sec:op}. Adapting that discussion to the $p = 6$ case, the local solution reads
\begin{equation}
  d   s^2 \sim t^{- 1 / 2} d   s^2_{\parallel} + t^{1 / 2} (d
    t^2 + d   s^2_{S^{8 - p}})~, \qquad e^{\phi} \sim t^{- 3 /
  4}~,  \label{eq:O6HoleLocalMetric} 
\end{equation}
with $t\equiv r-r_0$. Our aim in the following will be to find $\text{dS}_4$ solutions where singularities of the type {\eqref{eq:O6HoleLocalMetric}} are allowed. We will interpret these singularities as boundaries of the holes produced by
O6-planes. As we will see, solutions involving singularities \eqref{eq:O6HoleLocalMetric} will in general have free parameters, and a priori we may wonder if more restrictive boundary conditions can or should be imposed to fix them. This is similar to analogous issues that arise for the O8$_{-}$ solutions which are differed to section \ref{sec:o8o8}. 

We now look for dS$_4$ backgrounds of massive type IIA supergravity. We restrict ourselves to a co-homogeneity-one ansatz.\footnote{A similar Ansatz was used for de Sitter solutions in string theory and more generally with extra dimensions in \cite{EvaStrings}.} For simplicity, we also include an explicit $S^2$ factor in the metric, which will be the sphere transverse to the O6-plane. Thus our ansatz for the metric reads
\begin{equation}\label{eq:ans-met}
  d   s^2_{10} = e^{2 W} d   s^2_4 + e^{- 2 W} (e^{2 \lambda_3}
  d   s^2_{\kappa_3} + e^{2 Q} d   z^2 + e^{2 \lambda_2} d
    s^2_{S^2})~,
\end{equation}
with all the functions only depending on the coordinate $z$. The metric $ds^2_{\kappa_3}$ is an Einstein space with Einstein constant $\kappa_3$ and the function $Q$ parametrizes a gauge redundancy.

The most general fluxes compatible with the symmetries of the metric are
\begin{equation}\label{eq:ans-fluxes}
	H  =  h_{} d   z \wedge \text{vol}_2~, \hspace{.3in} F_2  =  f_2 \text{vol}_2~,\hspace{.3in}  F_4  =  f_{41} \text{vol}_3 \wedge d   z + f_{42} \text{vol}_4~, \hspace{.3in}  F_0  \neq 0~,
\end{equation}
where a priori $h, f_2, f_{41}$ and $f_{42}$ are all functions of the coordinate $z$.

\subsubsection{Equations of Motion} 
\label{ssub:o6-eom}

Away from sources, we can solve the Bianchi identities and equations of motion
for the fluxes by setting
\begin{equation}
  h_{} = f_2' / F_0    , \quad f_{42} = \text{cost}  ,
  \quad f_{41} = \frac{1}{F_0} e^{Q - 6 W - 2 \lambda_2 + 3 \lambda_3} (F_0
  c_1 - f_{42} f_2),
\end{equation}
where $c_1$ is an integration constant. The equations for the fluxes are then
completely satisfied up to the differential equation
\begin{equation}
  f_2'' = e^{2 (Q - 5 W + \phi)} (F_0 c_1 f_{42} + (e^{8 W} F_0^2 - f_{42}^2)
  f_2) + f_2' (Q' - 4 W' + 2 \lambda_2' - 3 \lambda_3' + 2 \phi')
  \label{eq:f2eom} .
\end{equation}
This local form of the equation of motion has to be supplemented with boundary
conditions for the fluxes, which we are going to discuss in the next section
where we focus on a specific choice for the sources.

Turning our attention to the Einstein and dilaton equations of motion, 
we obtain\footnote{For simplicity we display here only the equations of motion for the case with $F_4 = 0$.}
\begin{subequations}
  \allowdisplaybreaks
  \begin{align}
    &8 \Lambda  e^{2 q_0-4 W}&=&e^{4 W-4 \lambda _2}\left(-\frac{\left(f_2'\right)^2 }{F_0^2}+f_2^2 e^{2q_0-2W+2\phi}+F_0^2 e^{ 2q_0-6W+4\lambda_2+2\phi}\right)+\nonumber\\
    &&&-16 \lambda _2' \phi '-24 \lambda _3'
    \phi '+4 \left(\lambda _2'\right){}^2+12 \left(\lambda _3'\right)^2+24 \lambda _2' \lambda _3'+\nonumber\\
    &&&-6 \kappa _3 e^{2 q_0-2 \lambda _3}-4 e^{2 q_0-2 \lambda _2}+8 W' \phi '-16 \left(W'\right)^2+8 \left(\phi
    '\right)^2\\
    &16 W''&=&e^{4 W-4 \lambda _2}\left(-\frac{2 \left(f_2'\right)^2}{F_0^2}+6 f_2^2 e^{2q_0-2W+2\phi}+6 F_0^2 e^{2q_0-6W+4\lambda_2+2\phi}\right)+\nonumber\\
   &&&-32 \lambda _2' \phi '-48 \lambda_3' \phi '+8 \left(\lambda _2'\right){}^2+24 \left(\lambda _3'\right){}^2+48 \lambda _2'
   \lambda _3'+16   \left(\phi '\right)^2 +\nonumber\\
   &&&-12 \kappa _3 e^{2 q_0-2 \lambda _3}-8 e^{2 q_0-2 \lambda _2}+\frac{\delta _6
   \kappa ^2 \tau _6 e^{-2 \lambda _2+q_0+W+\phi }}{\pi }+\nonumber\\
   &&&+4 \delta _8 \kappa ^2 \tau _8
   e^{q_0-W+\phi }-32 \lambda _2' W'-48 \lambda _3' W'+48 W' \phi '-32 \left(W'\right)^2\\
   &8 \lambda _2''&=&e^{4 W-4 \lambda _2}\left(-5\frac{\left(f_2'\right)^2 }{F_0^2}+f_2^2 e^{2q_0-2W+2\phi}+5F_0^2 e^{ 2q_0-6W+4\lambda_2+2\phi}\right)+\nonumber\\
   &&&-24 \lambda _3' \phi '-12   \left(\lambda _2'\right){}^2+12 \left(\lambda _3'\right)^2-6 \kappa _3 e^{2 q_0-2 \lambda
   _3}+4 e^{2 q_0-2 \lambda _2}+\nonumber\\
   &&&+4 \delta _8 \kappa ^2 \tau _8 e^{q_0-W+\phi }+8 W' \phi '-16   \left(W'\right)^2+8 \left(\phi '\right)^2\\
   &8 \lambda _3''&=&e^{4 W-4 \lambda _2}\left(-\frac{\left(f_2'\right)^2 }{F_0^2}+5f_2^2 e^{2q_0-2W+2\phi}+5F_0^2 e^{ 2q_0-6W+4\lambda_2+2\phi}\right)+\nonumber\\
   &&&-16 \lambda _2' \phi '-8
   \lambda _3' \phi '+4 \left(\lambda _2'\right){}^2-12 \left(\lambda _3'\right){}^2+8 \lambda
   _2' \lambda _3'+\nonumber\\
   &&&+2 \kappa _3 e^{2 q_0-2 \lambda _3}-4 e^{2 q_0-2 \lambda _2}+\frac{\delta _6
   \kappa ^2 \tau _6 e^{-2 \lambda _2+q_0+W+\phi }}{\pi }+\nonumber\\
   &&&+4 \delta _8 \kappa ^2 \tau _8
   e^{q_0-W+\phi }+8 W' \phi '-16 \left(W'\right)^2+8 \left(\phi '\right)^2\\
   &4 \phi ''&=&e^{4 W-4 \lambda _2}\left(-2\frac{\left(f_2'\right)^2 }{F_0^2}+3f_2^2 e^{2q_0-2W+2\phi}+5F_0^2 e^{ 2q_0-6W+4\lambda_2+2\phi}\right)+\nonumber\\
   &&&-8 \lambda _2' \phi '-12
   \lambda _3' \phi '+\frac{3 \delta _6 \kappa ^2 \tau _6 e^{-2 \lambda _2+q_0+W+\phi }}{4 \pi
   }+5 \delta _8 \kappa ^2 \tau _8 e^{q_0-W+\phi }+8 \left(\phi '\right)^2 
     \end{align}
  \end{subequations}

  The first equation is a first order equation which will act as a constraint.
Each of the other four equations involves a second derivative of a different
function, and includes a $\delta$-function that accounts for the back-reaction of the physical sources. 
Since our
8-dimensional sources are O8$_+$-planes, which do not suffer from
strong-coupling ambiguities, the $\delta_8$ terms are well-defined. However,
the $\delta_6$ terms that appear in the equations of motion are just formal
devices since, as we have seen in the flat-space case, for O$p$-planes with $p <
7$ their support would be located inside the hole, and hence {\tmem{outside of the physical space-time described by supergravity}}.


\subsubsection{Flux Quantization} 
\label{ssub:flux}

From now on, we will focus on solutions with an O8$_+$ and an O6$_-$. The orientifold involution is generated by the operators
\begin{equation}
	\Omega_\mathrm{WS} \sigma_8 \, ,\qquad \Omega_\mathrm{WS} (-1)^\mathrm{F_\mathrm{L}} \sigma_6\,.
\end{equation}
$\Omega_\mathrm{WS}$ is the worldsheet parity; $F_\mathrm{L}$ is the left-moving fermion number operator; $\sigma_8:\, z\mapsto -z$, whose fixed locus is at $z=0$, the O8-plane; and $\sigma_6:\, (\theta,\phi)\mapsto (\pi-\theta,-\phi) $ is the antipodal map on the $S^2$, whose fixed locus is at the locus where the $S^2$ shrinks, the O6-planes. The difference between O$p_\pm$ is in general defined via the orientifold action on the Chan--Paton variables, but in cases without D-branes such as this one it can also be defined as a sign appearing for non-orientable maps at the O$p_+$, as explained for example in \cite[Sec.~6.2]{witten-without}. Early models with O-planes of different signs appearing simultaneously appeared for example in \cite{bianchi-pradisi-sagnotti,angelantonj-bianchi-pradisi-sagnotti-stanev,witten-without}.

The $\sigma_z$ involution means that the physics in $z<0$ is just a replica of the physics in $z>0$; so, as is often done, we will consider only the latter half. Then, $z$ starts from an O8$_+$ plane sitting at $z = 0$ and ends at the hole of an O6-plane at $z = z_0$. 

We will restrict our attention to the case
\begin{equation}\label{eq:F4=0}
	F_4 = 0\,.
\end{equation}
The presence of an O8$_+$ will make the flux $F_0$ jump according to its
Bianchi identity:
\begin{equation}
  \Delta F_0 = - \kappa^2 \tau_8\, \label{eq:F0jump} .
\end{equation}
Since $F_0$ is odd across an O8-plane, we have $\Delta F_0 = 2   F_0 |_{z \rightarrow 0^+}$. Combining the two equations we get in our conventions
\begin{equation}
    F_0 |_{z \rightarrow 0^+} = \frac{n_0^+}{2 \pi} \, ,\qquad
  n_0^+ = - 4\,. \label{eq:fluxCond1}
\end{equation}
The behavior of $F_2$ on the O8 plane requires some care. Away from O6/D6 and NS5/ONS5 we have to satisfy the Bianchi identities
\begin{equation}\label{eq:bianchi}
  d   F_2 = F_0 H\,, \qquad d   H = 0\, .
\end{equation}
In particular $H$ does not have to jump. Since on top of an O8$_+$ plane $F_0$ jumps according to {\eqref{eq:F0jump}}, then $d   F_2$ has to jump.

The O6 at $z = z_0$ is not defined through a $\delta$-function, since the $\delta_6$ is outside of the space-time, but through the boundary condition
\begin{equation}
  f_2 (z_0) = 1. \label{eq:fluxCond2}
\end{equation}
This choice fixes the flux quantization for $F_2$.

Finally, we have to impose flux quantization for $H$:
\begin{equation}
  \int_{M_3} H = (2 \pi)^2 N . \label{eq:fluxQuantH}
\end{equation}
To do so, we integrate the Bianchi equation on half of the internal space,
\begin{equation}
  \int_{\frac{M_3}{2}} d   F_2 = \int_{\frac{M_3}{2}} F_0 H,
\end{equation}
and we use {\eqref{eq:fluxQuantH}} and the fact that $H$ as a form is even across the O8$_+$ to obtain
\begin{equation}
  4 \pi (f_2 (z_0) - f_2 (0)) = F_0^+ \frac{1}{2} (2 \pi)^2 N .
\end{equation}
By writing $F_0 \equiv \frac{n_0}{2 \pi}$ and using {\eqref{eq:fluxCond2}}, we get
\begin{equation}
  f_2 (0) = 1 - \frac{n_0^+ N}{4}  \label{eq:fluxCond3}
\end{equation}
where for a simple O8+ (i.e. without D8s on top of it) $n_0^+ = - 4$.

Summing up, for a solution of the type O8$_+$-O6 we have to impose the
conditions {\eqref{eq:fluxCond1}}, {\eqref{eq:fluxCond2}} and
{\eqref{eq:fluxCond3}}, which account for the flux quantization of $F_0$,
$F_2$ and $H$.


\subsubsection{O8$_+$ Boundary Conditions and the
Cosmological Constant} 
\label{ssub:o6-bc}

By integrating the equations of motion across the O8$_+$plane at $z = 0$, we
obtain the boundary conditions
\begin{equation}
  \lambda_2' = \lambda_3' = - \frac{1}{2} F_0 e^{q_0 - W + \phi}, \qquad W' =
  - \frac{1}{4} F_0 e^{q_0 - W + \phi}, \qquad \phi' = - \frac{5}{4} F_0
  e^{q_0 - W + \phi}, \quad \quad \text{at } z = 0. \label{eq:O8bcs}
\end{equation}
By plugging these conditions into the constraint equation we get
\begin{eqnarray}
  \Lambda & = & \frac{1}{8} f_2^2 e^{- 4 \lambda_2 + 6 W + 2 \phi} -
  \frac{3}{4} \kappa_3 e^{4 W - 2 \lambda_3} - \frac{1}{2} e^{4 W - 2
  \lambda_2} - \frac{(f_2')^2 e^{- 4 \lambda_2 - 2 q_0 + 8 W}}{8 F_0^2} + \\
  & + & \frac{c_1 f_2 f_{42} e^{- 4 \lambda_2 - 2 W + 2 \phi}}{4 F_0} -
  \frac{1}{8} c_1^2 e^{- 4 \lambda_2 - 2 W + 2 \phi} - \frac{f_2^2 f_{42}^2
  e^{- 4 \lambda_2 - 2 W + 2 \phi}}{8 F_0^2} - \frac{1}{8} f_{42}^2 e^{2 \phi
  - 6 W}  \nonumber\\
  &  & \text{at } z = 0 \nonumber
\end{eqnarray}
where the second line vanishes for $F_4$ = 0.  In particular we notice that with $\kappa_3$ negative enough we can obtain a positive cosmological constant.



\subsection{An Analytic AdS Starting Point} 
\label{sub:ads7}

A notable class of already known solutions that fits in our Ansatz (\ref{eq:ans-met}), (\ref{eq:ans-fluxes}), (\ref{eq:F4=0}) can be obtained from the AdS$_7$ solutions in \cite{afrt,10letter,cremonesi-t} by replacing simply
\begin{equation}\label{eq:74}
	\mathrm{AdS}_7 \to \mathrm{AdS}_4\times H_3\,,
\end{equation}
where $H_3$ is a compact hyperbolic space with the same Einstein constant as AdS$_4$, $\kappa_3 = \Lambda$, and doing nothing else. At the level of the equations of motion this replacement has no impact. (At the level of supersymmetry equations it does make a difference; there is a procedure to generate supersymmetric $\mathrm{AdS}_4\times H_3$ solutions from AdS$_7$ supersymmetric ones, but it is more complicated and involves also changing the internal metric and fluxes in a certain way \cite{rota-t}, in the spirit of \cite{maldacena-nunez}.)

For the solutions obtained this way, the local form of the metric functions is given by
\begin{equation}
  e^{2 W} = e^{\lambda_3}= \sqrt{2} \pi \sqrt{- \frac{\alpha}{\ddot{\alpha}}}\,, \qquad e^{2 \lambda_2} =
  \frac{2 \pi^2 X^{5 / 2} \alpha^2}{X^5 \dot{\alpha}^2 - 2 \alpha \ddot{\alpha}}\,,\qquad q_0 = 2 \pi^2 X^{- 5 / 2}\,, \qquad \Lambda = - \frac{2 + X^5}{4  X^{5 / 2}} \,;
\end{equation}
so in particular the metric is given by
\begin{equation}\label{eq:ads7met}
  \frac{1}{\sqrt{2} \pi} d   s^2_{10} = \sqrt{- \frac{\alpha}{\ddot{\alpha}}} (d
    s^{_2}_{\text{AdS}_4} + d   s^2_{H_3}) + \sqrt{-
  \frac{\ddot{\alpha}}{\alpha}} X^{- 5 / 2} \left( d   z^2 + \frac{\alpha^2}{\dot{\alpha}^2
  - 2 X^{- 5} \alpha \ddot{\alpha}} \right) \,.
\end{equation}
Reality and positivity of the metric are achieved if 
\begin{equation}\label{eq:ge0}
	\alpha\ge 0 \, ,\qquad - \ddot \alpha>0\,.
\end{equation}
$F_2$ and the dilaton are obtained from
\begin{equation}
  e^{\phi} = X^{5 / 4} \frac{2^{5 / 4} 3^4 \pi^{5 / 2} \left( -
  \frac{\alpha}{\ddot{\alpha}} \right)^{3 / 4}}{\sqrt{X^5  \dot{\alpha}^2 - 2 \alpha \ddot{\alpha}}},
  \qquad f_2 = \frac{\ddot{\alpha}}{2 3^4 \pi^2} + \frac{F_0 \pi X^5 \alpha \dot{\alpha}}{X^5
  \dot{\alpha}^2 - 2 a \ddot{\alpha}} \,.
\end{equation}
(Since we are interested in the case where $F_0\neq 0$, we can take $B=\frac{F_2}{F_0}$, which automatically solves (\ref{eq:bianchi}).) Both values of the constant
\begin{equation}
	X=1 \, ,\qquad X=2^{1/5}
\end{equation}
lead to a solution of the equations of motion. In AdS$_7$, the supersymmetric solution is obtained for $X$=1, while the non-supersymmetric one is obtained for $X = 2^{1 / 5}$ \cite{passias-rota-t}. After our replacement (\ref{eq:74}), both cases are non-supersymmetric. However, in what follows we will focus on the $X=2^{1/5}$ case.

The equations of motion force $\alpha$ to be a piece-wise degree 3 polynomial that
has to satisfy
\begin{equation}
  \dddot{\alpha} = - 162 \pi^3 F_0\,.
\end{equation}
If 8-dimensional sources are present, $F_0$ changes accordingly to its Bianchi identity, and $\dddot{\alpha}$ jumps. Nevertheless one can impose that the metric and fields are continuous.

Different sources are then chosen by specifying the correct boundary
conditions for $\alpha$, which has three free parameters. We highlight the following: 
\begin{itemize}
	\item A D6 is obtained by imposing $\alpha\to 0$. 
	\item The boundary of an O6 hole is obtained with $\ddot \alpha\to0$. 
	\item An O8 requires $\dot\alpha \to 0$.\footnote{The ``diverging dilaton'' type, which is only possible for an O8$_-$, is obtained by imposing that $\ddot\alpha\to 0$ at the same time. We will not need this here.}
\end{itemize}

To obtain an O8$_+$-O6 solution we impose the conditions
\begin{equation}\label{eq:O8O6-oldbc}
  \dot{\alpha} (0) = 0, \qquad \ddot{\alpha} (z_0) = 0, \qquad f_2 (z_0) = 1,
  \qquad f_2 (0) = \frac k2 = 1 - \frac{n_0^+ N}{4} \quad  \,,
\end{equation}
where both $k$ and $N$ should be integers. The first two conditions in (\ref{eq:O8O6-oldbc}) come from imposing that at $0$ and $z_0$ the solution has the correct local behavior for an O8$_+$ and for the boundary of an O6$_-$ boundary respectively, as in the list of possibilities above. The third condition fixes the charge of the O6$_-$ (which is $-2$ that of a D6). Finally, the fourth takes care simultaneously of flux quantization for $F_2$ at $z=0$, and of flux quantization for $H$, whose flux integral (over the whole space, from $-z_0$ to $z_0$) is $N$. The two are related by integrating the Bianchi. 
If we have a simple O8$_+$ at $z = 0$, then $n_0^+ = - 4$, and we are only left
with the freedom of choosing the integer $N$. Moreover, in this gauge $z_0$
depends on $N$ as
\begin{equation}
	z_0 =  -\frac N2 + \frac2{n_0^+}=- \frac{N + 1}{2} \label{eq:z0O6hole},
\end{equation}
and the requirement that $z_0 > 0$ forces $N < - 1$. Explicitly, the solution reads
\begin{equation}\label{eq:aO8O6}
	\alpha= \frac{27}{32}\pi^2 \left(k^2 (k+12 N) + 48 k z^2 + 64 z^3\right)\,;
\end{equation}
recall $k=2(N+1)$.

This solution can be checked using holography: the  $a_\mathrm{Weyl}$ anomaly can be computed using both field theory and holography as in \cite{cordova-dumitrescu-intriligator-a6, cremonesi-t}, getting the same result $\frac{16}7 {2}{15}N^3 k^2$ in the limit where $N \gg1$ \cite{apruzzi-fazzi}.

Curiously, for all AdS$_7$ solutions with an O6 there is the possibility of analytically continuing past the boundary of the O6 hole. This requires going past the point where $\ddot \alpha=0$, to a region where $\ddot \alpha>0$, violating (\ref{eq:ge0}) and making (\ref{eq:ads7met}) imaginary, exactly as for the hole in (\ref{eq:Op}), (\ref{eq:HOp}). The center of this hole is obtained at a point where $\alpha\to0$, where there is a formal singularity $z=z_\mathrm{O}$ similar to the one of a D6 but where some functions have opposite signs. For the solutions (\ref{eq:aO8O6}) above, this happens at a point $z_\mathrm{O}\sim -N/2$.\footnote{In (\ref{eq:O8O6-oldbc}), we have imposed flux quantization $\int H = 4\pi^2 N$ over the physical region alone; one might wonder what happens if one extends the integral over the hole as well. Formally this is achieved by imposing the same conditions at $z_\mathrm{O}$ as one would impose for a D6 stack, but taking $n_\mathrm{D6}=-2$. The solution obtained in this way is $\tilde \alpha= \alpha -\frac{27}4 \pi^2$, so the difference between the two is small when $N$ is large.} In general, it is not clear that this procedure of continuing classical supergravity solutions beyond the hole (where they are not valid) has any physical meaning.  In our numerical dS$_4$ solutions below we are not able to completely reproduce this continuation.


\subsection{Numerical Solutions} 
\label{sub:num}

In this section, we show that it is possible to find dS$_4$ solutions of the
type O8-O6, if the O6 is identified by the behavior of the metric and the
dilaton near its hole, as in the flat space case
{\eqref{eq:O6HoleLocalMetric}}. Imposing these conditions, we are able to
explicitly build the numerical solutions. We find a three-parameter family of
solutions labeled by the boundary data for the unconstrained metric functions
at the boundary of the hole. 

We start by building the local solution near an O8$_+$-plane at $z = 0$. By
imposing the boundary conditions {\eqref{eq:O8bcs}} we obtain the expressions
\begin{eqnarray}
  e^{- 4 W} & = & 1 + \frac{F_0 e^{q_0} z}{a_1^{3 / 4}} + \frac{1}{2} e^{2
  q_0} z^2  \left( - \frac{f_{20}^2}{a_1^{3 / 2} a_2^2} - 4 \Lambda \right) +
  O (z^3)~, \nonumber\\
  e^{- \frac{4}{3} \phi} & = & a_1 + \frac{5}{3}  \sqrt[4]{a_1} F_0 e^{q_0} z
  + \frac{z^2  \left( \frac{6 a_1^{3 / 2} b^2}{F_0^2} + e^{2 q_0}  (10 a_2^2
  F_0^2 - 9 f_{20}^2) \right)}{18 \sqrt{a_1} a_2^2} + O (z^3)~,\nonumber \\
  e^{- 2 \lambda_3} & = & 1 + \frac{F_0 e^{q_0} z}{a_1^{3 / 4}} + \frac{z^2 
  \left( 2 e^{2 q_0}  \left( a_2  (a_2 \Lambda + 2) - \frac{2 f_{20}^2}{a_1^{3
  / 2}} \right) + \frac{b^2}{F_0^2} \right)}{6 a_2^2} + O (z^3)~, \\
  e^{2 \lambda_2} & = & a_2 - \frac{a_2 F_0 e^{q_0} z}{a_1^{3 / 4}} + z^2 
  \left( e^{2 q_0}  \left( \frac{a_2 F_0^2}{a_1^{3 / 2}} + a_2 \Lambda + 1
  \right) - \frac{b^2}{2 a_2 F_0^2} \right) + O (z^3)~, \nonumber \\
  f_2 & = & f_{20} + bz + \frac{F_0 e^{q_0} z^2  (f_{20} F_0 e^{q_0} - a_1^{3
  / 4} b)}{2 a_1^{3 / 2}} + O (z^3)~. \nonumber
\end{eqnarray}
Some comments are in order.
\begin{itemize}
  \item Since we decided to keep $\Lambda$ and $\kappa_3$ as continuous
  parameters, we fixed the redundancy in the parametrization of the metric by
  setting $e^{- 4 W}$ and $e^{- 2 \lambda_3}$ equal to 1 on top of the O8$_+$.
  
  \item $q_0$ here is just a gauge redundancy and we can use it to rescale the length of the interval.
  
  \item $f_{20}$ and $F_0$ are discrete parameters depending on $N$ and $n_0$
  as in {\eqref{eq:fluxCond1}} and {\eqref{eq:fluxCond3}}:
  \begin{equation}
    f_{20} = 1 - \frac{n_0^+ N}{4}~, \qquad F_0 = \frac{n_0^+}{2 \pi} ~.
  \end{equation}
  For a simple O8$_+$ without D8-branes on top of it we have $n_0^+ = - 4$.
  
  \item $b$ is not a free parameter.  From the first order equation we find:
  \begin{equation}
    b = \pm F_0 e^{q_0} \sqrt{\frac{f_{20}^2}{a_1^{3 / 2}} - 2 a_2  (3 a_2
    \kappa_3 + 4 a_2 \Lambda + 2)}  \label{eq:equationb} ~.
  \end{equation}
  The two roots correspond to the 2 possible choices for the sign of $f_2'
  (0)$. We find that only the positive root gives the solutions we are
  interested in. Moreover, notice that in order to have real solutions the
  expression inside the square root has to be non-negative. This gives a
  inequality on the initial parameters of a physically acceptable solution.
\end{itemize}
To summarize, the local solution near the O8$_+$-plane depends on four continuous
parameters $a_1, a_2, \kappa_3$ and $\Lambda$ and two discrete ones, $N$ and
$n_0^+$. These parameters have to be chosen such that $b$ defined in
{\eqref{eq:equationb}} is real. To hit an O6$_-$ we need to find a point where
$f_2 = 1$, requiring one fine-tuning.

\begin{figure}[h]
  \centering
  \resizebox{234pt}{151pt}{\includegraphics{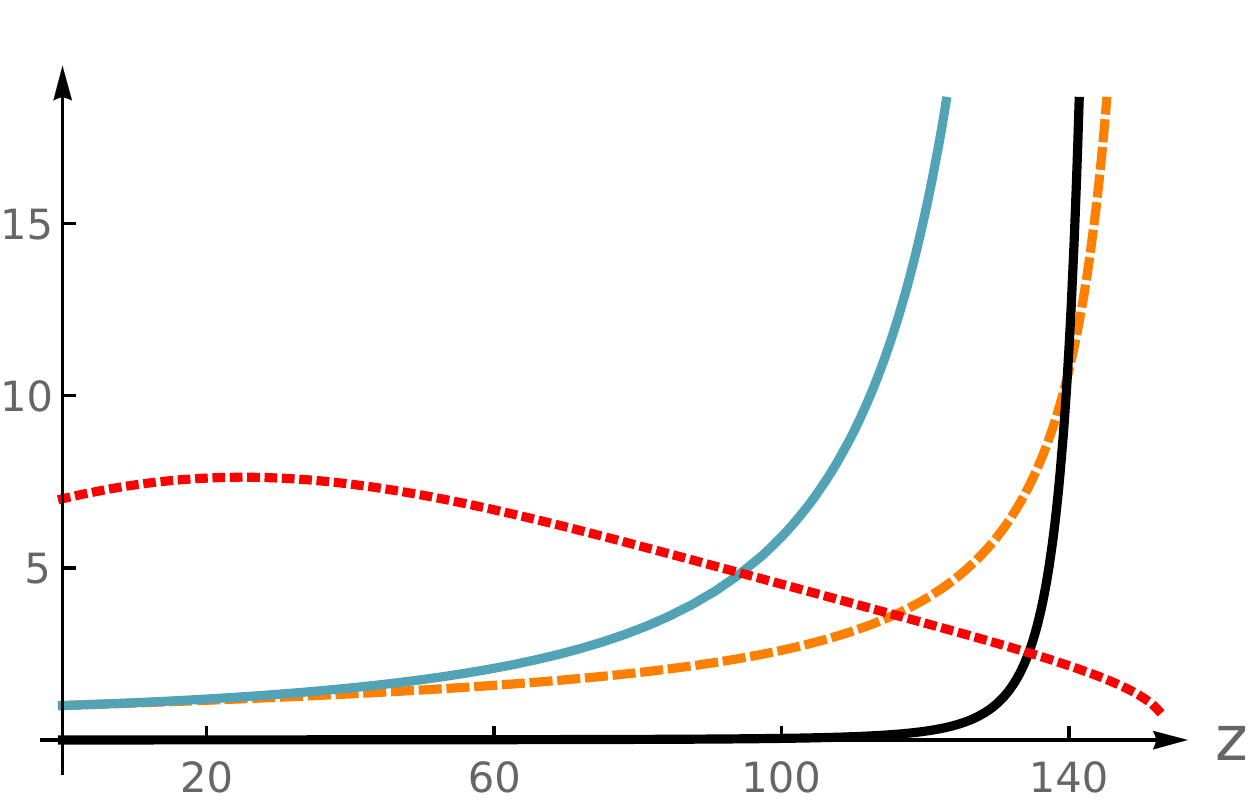}}
  \caption{A numerical dS$_4$ solution with $\Lambda = 2.7 \times 10^{- 3}$
  and $\kappa_3 = - 2.1 \times 10^{- 1}$. It starts from an O8$_+$-plane (on
  the left) and ends at the boundary of the hole produced by an O6$_-$-plane
  (on the right). The functions are $e^{4 W}$(turquoise), $e^{4 \phi}$(black),
  $e^{2 \lambda_3}$(orange, dashed) and $e^{2 \lambda_2}$(red, dashed). On the
  right, the functions behave as in equation
  {\eqref{eq:numericalO6behavior}}.\label{fig:O6UpToHole}}
\end{figure}

We now take an AdS$_4 \times H_3$ solution at large $N$, i.e.~weakly-curved and weakly-coupled, 
and we slowly increase $\Lambda$ making it
positive. Correspondingly, we have to tune the parameters ($\kappa_3, a_1,
a_2$) in order to reach a point $z_0$ where, defining $t \equiv |z - z_0|$, the
functions behave as
\begin{equation}
  f_2 (z_0) = 1, \qquad e^{\lambda_2} \sim \text{const}, \qquad e^W \sim t^{-
  \frac{1}{4}}, \qquad e^{\lambda_3} \sim t^{- \frac{1}{2}}, \qquad e^{\phi}
  \sim t^{- \frac{3}{4}} \quad \label{eq:numericalO6behavior} .
\end{equation}
Near such a point, the metric, the dilaton, and the fluxes have the same local
expression as in {\eqref{eq:O6HoleLocalMetric}}. As in that case, the supergravity
approximation breaks down near the boundary of hole,
since the dilaton starts growing and eventually diverges. Figure
\ref{fig:O6UpToHole} shows a typical solution with this behavior.

We have thus succeeded in obtaining dS$_4$ solutions with an O8$_+$ at $z=0$ and an O6$_-$ whose boundary lies at $z=z_0$. Near this boundary the dilaton and string coupling diverge, but they do in the same way as for the O6 in flat space and for the O6 in the AdS$_7$ solutions in section \ref{sub:ads7}.

As we have mentioned earlier, in this case it is unclear how one should check whether the $\delta$'s in the equations of motion are correctly reproduced, since their support now lies outside the physical space, in the ``hole'' region where the metric is purely imaginary. Moreover, the integration by parts argument that was in \cite{cribiori-junghans} to criticize the O8$_+$--O8$_-$ solutions of \cite{cordova-deluca-t-ds4} in this case doesn't apply: if we consider \cite[(A.7)]{gautason-junghans-zagermann}, there is no single choice of $c$ that makes both the O8$_+$ and the O6$_-$ source terms disappear.\footnote{ A speculative way to test the $\delta$ source term of the O6 might be to formally try to continue the solutions inside the unphysical ``hole''. It is unclear how physical this is, but it does work for the analytic AdS$_7$ solutions with O6-planes, as we noticed at the end of section \ref{sub:ads7}. We attempted this by continuing the numerical evolution past the O6 boundary. This requires fine-tuning on the initial conditions, and  we did not manage to obtain the full formal hole behavior.  For instance, $H \equiv e^{- 4 W}$ is not diverging and $e^{\lambda_i}$ and $e^{\phi}$ tend to zero with with unusual power laws. We thank N. Cribiori and D. Junghans for correspondence about this issue.
}

Finally, we comment on the physical value of the cosmological constant. At this
stage, $\Lambda$ is only a parameter. Its physical value in units of the four
dimensional Planck mass is obtained as
\begin{equation}
  \Lambda_{\text{phys}} = \frac{\Lambda}{M_p^2}, \qquad \text{with} \quad
  M_P^2 = \kappa^2 \text{Vol}_2 \text{Vol}_3 \int_{} d   z e^{Q - 4 W +
  2 \lambda_2 + 3 \lambda_3 - 2 \phi} .
\end{equation}
Performing this integral\footnote{There is almost no difference in stopping
the integral on the boundary of the hole or on top of the O6 since the
contribution of the hole is very small.} for the solution in Figure
\ref{fig:O6UpToHole} we obtain
\begin{equation}
  \Lambda_{\text{phys}} \sim \frac{1}{\kappa^2} \text{Vol}_3 10^{- 9} .
\end{equation}
This small number is expected from the AdS$_7$ solutions, where at large $N$
the Planck's mass scales as $N^5$ and $\Lambda$ remains constant.



\section{A Discussion of the O8$_+$--O8$_-$ Solutions} 
\label{sec:o8o8}

In this section we discuss the original O8$_+$--O8$_-$ de Sitter solutions of \cite{cordova-deluca-t-ds4} in detail.  In particular our analysis is focused on the behavior near the O8$_-$.

In \cite{cribiori-junghans}, it was claimed that such solutions would be impossible quite generally, based on a certain integration by parts applied to the supergravity equations of motion. They explained the disagreement with \cite{cordova-deluca-t-ds4} by claiming that the $z=z_0$ singularity, while displaying the same behavior as an O8$_-$ at leading order in $|z-z_0|$, was crucially different from it in subleading behavior. In this section we will investigate this claim. 

In appendix \ref{bdsusy} we also exhibit several previously known \cite{passias-prins-t} analytic supersymmetric AdS$_{4}$ solutions with O$8_{-}$ planes and the same subleading behavior as our de Sitter solutions.

\subsection{The Solutions in Detail} 
\label{sub:recap}

Let us then first look at the equations of motion from the de Sitter solutions of \cite{cordova-deluca-t-ds4} for the functions $W$, $\phi$, $\lambda$ defined in the ansatz \eqref{eq:metriceasy}.  Since the contentious point is the treatment of O-planes, we include the source terms:
\begin{subequations}\label{eq:eom}
\begin{align}
	& W'' +W'(5 \lambda - 2 \phi)'-\frac14 F_0^2 e^{2(\phi-W)}-\Lambda e^{-4W} =  \frac1\pi	 e^{\phi-W} \sigma\,;
	\label{eq:eom1}\\
	& (W+2 \phi - 5 \lambda)''+W'(5 \lambda+2 \phi)'-8 (W')^2 -5 (\lambda')^2 +\frac14 F_0^2 e^{2(\phi-W)}=  \frac1\pi e^{\phi-W} \sigma \,; \label{eq:eom2}\\
	& (W-\lambda)''+(W-\lambda)'(5 \lambda - 2 \phi)' + \kappa e^{-2 \lambda} +\frac14 F_0^2 e^{2(\phi-W)}= -\frac1\pi e^{\phi-W} \sigma\,; \label{eq:eom3}\\
	 & 4 (W')^2-10 (\lambda')^2 -2 (\phi')^2 +2\phi'(5 \lambda-W)'+ 2 e^{-4W} \Lambda + \frac52 \kappa_5 e^{-2 \lambda}-\frac14 F_0^2 e^{2(\phi-W)} = 0\,
\end{align}											
\end{subequations}
where 
\begin{equation}
	\sigma \equiv \delta(z)- \delta(z-z_0)\,.
\end{equation}
is the sum of the O8$_\pm$ localized contribution.

To focus on the sources, we notice that the second-order equations (\ref{eq:eom1})--(\ref{eq:eom3}) are linear combinations of equations of the form 
\begin{equation}\label{eq:eom-dots}
	e^{W-\phi}\partial_z^2 f_i = \pm \frac1\pi \delta +\ldots \,\qquad (\text{O8}_{\pm}) \,.
\end{equation}
From now on we omit the argument of the $\delta$ when it should be obvious from the context. The $\ldots$ denote terms that are not important near the sources, and
\begin{equation}
	f_i \equiv \left\{W,\,\frac15\phi,\, \frac12 \lambda\right\}\,.
\end{equation}

Near the O8$_+$, the functions are finite; assuming they are also continuous, the treatment of the $\delta$ terms is standard. One can for example take the integral of the sources on a small interval $[-\epsilon,\epsilon]$; in the $\epsilon\to 0$ limit, the only contributions come from the $\delta$ terms and from the discontinuities in the second derivatives. Alternatively, we can directly use the fact that the weak derivative of a discontinuous function $h$ includes a $\delta$ term proportional to the discontinuity; schematically
\begin{equation}\label{eq:weak-der}
	\partial_z h= h' + (\Delta h) \delta\,.
\end{equation}
 Either way, we obtain
\begin{equation}\label{eq:bc-O+}
	e^{W-\phi} f_i'|_{z \to 0^+} = \frac1{2\pi}\, \qquad (\text{O8}_+)\,.
\end{equation} 

The O8$_-$ is more subtle; near its position $z_0$, the leading behavior in $z-z_0$ of the $f_i$ is 
\begin{equation}\label{eq:efi-O8-}
	e^{f_i}= c_i |z-z_0|^{-1/4}+ O(|z-z_0|^{3/4})\,;
\end{equation}
so the $f'_i = \frac1{|z-z_0|} + {\rm reg}.$, and the $c_i$ are such that $e^{W-\phi}= |z-z_0|+O(|z-z_0|)$.

The logic we just used to obtain (\ref{eq:bc-O+}) now does not apply straightforwardly. For example, if we take the integral of (\ref{eq:eom-dots}) on $[-\epsilon,\epsilon]$, we have the integral of $e^{W-\phi}\partial_z^2 f_i$, which is not a total derivative. If one tries to use (\ref{eq:weak-der}), one is confronted with the derivative of functions $f_i$ which are not simply discontinuous but in fact divergent. One could alternatively multiply (\ref{eq:eom-dots}) by $e^{\phi-W}$, and then integrate it on $[-\epsilon,\epsilon]$. Now the left-hand side $\partial_z^2 f_i$ is a total derivative, but the right-hand side reads
\begin{equation}\label{eq:1/z-delta}
	e^{\phi-W} \delta\sim \frac1{|z-z_0|} \delta(z-z_0)
\end{equation}
which is a product of distributions, of unclear interpretation.

In fact this formal trouble hides an even deeper problem: since the dilaton and curvature are diverging at the O8$_-$, the supergravity approximation is breaking down there, and we shouldn't use the equations of motion (\ref{eq:eom}) in the first place. Our numerical solution cannot be trusted there, and trying to understand its formal properties is not physically meaningful. The reason we identified our divergence with that of an O8$_-$ was that its leading-order divergence in $(z-z_0)$ behaves exactly like an O8$_-$ solution in flat space.\footnote{\label{foot:O8-}The O8$_-$ in flat space has metric (\ref{eq:Op}), (\ref{eq:HO8-flat}). As we saw there, the particular case of this solution where $a=0$ has diverging dilaton and curvature; this is the solution that we will refer to as ``flat-space O8$_-$'' from now on.} That solution also has the problem that supergravity breaks down in its vicinity, but it is believed to be modified in fully-fledged string theory, whatever its equations of motion may be, but to still exist. 

More precisely, both the flat-space O8$_-$ and our solution satisfy 
\begin{equation}\label{eq:perm}
	e^{W-\phi} f_i'|_{z \to z_0^+} = -\frac1{2\pi}\, \qquad (\text{O8}_-)\,
\end{equation}
as one can see from (\ref{eq:efi-O8-}).
This is very similar to (\ref{eq:perm}); for an O8$_-$ where the dilaton and curvature remains small ($a\gg1$, in the language of footnote \ref{foot:O8-} and (\ref{eq:HO8-flat})), this could be motivated by the same arguments that took us to (\ref{eq:perm}). 

This similarity between our singularity and the flat-space O8$_-$ motivates the hope that our solution also exists in fully-fledged string theory. Notice that the solution has a supergravity ``modulus'' $c$, generated by the rescaling
\begin{equation}\label{eq:resc}
	g_{MN}\to e^{2c} g_{MN} \, ,\qquad \phi\to \phi-c\,.
\end{equation}
This is a symmetry of the supergravity equations of motion, but not of full string theory. We thus expect the solution to exist in string theory only for a particular value of $c$. (We will see later that it is not clear whether $c$ can in fact be considered a field in an effective four-dimensional description.) However, as stated in \cite{cordova-deluca-t-ds4}, all this is unproven until one somehow finds a way to evaluate the stringy corrections: there might be no value of $c$ for which the solution works in full string theory.

Meanwhile, \cite{cribiori-junghans} examined the same solution by using the supergravity equations of motion even in the region where supergravity breaks down. \emph{This should not be done}, and that the fate of the solution cannot possibly be decided this way. It is true that we ourselves used supergravity, but we just looked at the leading behavior of the fields in $|z-z_0|.$ In particular this allows us to read off the correct charge of the O8$_{-}$. However, it is not clear that applying the uncorrected supergravity equations of motion to the subleading behavior as in \cite{cribiori-junghans} makes any physical sense.  For the rest of this section we will consider this issue further.


\subsection{Various Versions of the Boundary Conditions} 
\label{sub:bc}

As we just saw, the solutions in \cite{cordova-deluca-t-ds4} satisfy (\ref{eq:perm}). However,
(\ref{eq:perm}) would seem to suffer from an ambiguity. For example one might want to rewrite it as
\begin{equation}\label{eq:restr}
		f_i'|_{z \to z_0^+} = - \frac1{2\pi} e^{W-\phi}|_{z \to z_0^+} \,.
\end{equation}
From (\ref{eq:efi-O8-}) we now see that both sides have a simple pole; so the equation is of the form 
\begin{equation}\label{eq:deLR}
	 \frac{d_i^\mathrm{L}}{|z-z_0|} + e_i^\mathrm{L} = \frac{d_i^\mathrm{R}}{|z-z_0|} + e_i^\mathrm{R} \,.
\end{equation}
Equality of the leading term, $d_i^\mathrm{L}= d_i^\mathrm{R}$, is equivalent to (\ref{eq:perm}); but if we also impose equality of the subleading coefficient $e_i^\mathrm{L}= e_i^\mathrm{R}$ we have a more restrictive boundary condition than (\ref{eq:perm}). By multiplying (\ref{eq:perm}) by a diverging function, we have made one more coefficient of its Taylor expansion emerge; this new coefficient has effectively created an extra boundary condition. Thus (\ref{eq:restr}) is \emph{not} in fact equivalent to (\ref{eq:perm}); we will call them the \emph{restrictive} and \emph{permissive} boundary conditions, respectively.

One could be even more perverse and multiply (\ref{eq:perm}) by an even more diverging function, such as $e^{2(W-\phi)}$. The two sides of the equation would now be of the form
\begin{equation}
	\frac{d_i^\mathrm{L,R}}{|z-z_0|^2} + \frac{e_i^\mathrm{L,R}}{|z-z_0|}+ f_i^\mathrm{L,R}\,,
\end{equation}
 thus creating the need to equate even the coefficients $f_i^\mathrm{L,R}$. While of course this does not appear particularly natural, one does not see any a priori reason to consider this option to be any less valid than (\ref{eq:perm}) or (\ref{eq:restr}). Clearly we need a better understanding of this ambiguity.

The restrictive boundary condition (\ref{eq:restr}) is effectively the one used in \cite{cribiori-junghans}; it is not satisfied by the singularity we identified as O8$_-$ in \cite{cordova-deluca-t-ds4}, and this is what creates the apparent contradiction between the two papers. 

More precisely, \cite{cribiori-junghans} applies (\ref{eq:restr}) to the difference
\begin{equation}
	(f_1-f_2)'=W'-\frac15 \phi'\,,
\end{equation}
which from (\ref{eq:deLR}) we see to be of the form $\frac{ \Delta d_{12}}{|z-z_0|}+ \Delta e_{12}=0$ for some constants $\Delta d_{12}$ and $\Delta e_{12}$.  The restrictive (\ref{eq:restr}) imposes $(f_1-f_2)'=0$, so both $\Delta d_{12}= \Delta e_{12}=0$; the permissive (\ref{eq:perm}) imposes $e^{W-\phi}(f_1-f_2)'=0$, so only $\Delta d_{12}=0$. Our solution in \cite{cordova-deluca-t-ds4} satisfies $\Delta d_{12}=0$: it cancels the pole in $(f_1-f_2)'$, but not the constant. 

In other words, \cite{cribiori-junghans} can be read as a complaint that in $(f_1-f_2)'$ we only made sure that the pole $\frac{ \Delta d_{12}}{|z-z_0|}$ vanished, and not the constant term $\Delta e_{12}$. Indeed they went further: with an integration by parts, they showed that no solution with only O8-planes exists such that this constant term vanishes.\footnote{This is based on \cite[(A.7)]{gautason-junghans-zagermann} taking $c=\frac25$. This is the same linear combination of equations of motion, but now including source terms, as the one used in \cite[Sec.~6.3]{maldacena-nunez} to extend the dS no-go to solutions with $F_0$.} So the issue is really reduced to whether we should use their restrictive boundary conditions (\ref{eq:restr}) for $(f_1-f_2)'= W'-\frac15 \phi'$, or our permissive ones.\footnote{A related point is that in \cite{cordova-deluca-t-ds4} we used a different set of variables, trading $\lambda$ for $\alpha\equiv e^{5 \lambda - 2 \phi}$. This makes it natural to impose $\alpha'=0$.  However, using \eqref{eq:efi-O8-} one can see that this imposes the restrictive boundary condition on the difference $f_2-f_3$.  If we relax this to the permissive boundary condition, consistent with our treatment of $f_1-f_2,$ leads to additional moduli and new solutions \cite{deluca-ds}.} Let us stress once again, however, that one really should be looking at the full string theory equations of motion.


\subsection{Action Variation} 
\label{sub:variation}

We will now see that the difference between permissive and restrictive boundary conditions (\ref{eq:perm}), (\ref{eq:restr}) can be traced back to how one varies the action: namely, to what space the variations $\delta f_i$ are taken to belong. We will illustrate this point by focusing on the dilaton's equation of motion. 

The relevant terms of the action in the string frame are
\begin{align}
	\label{eq:S0}	S^0 &= \frac1{\kappa^2} \int_{M_{10}}d^{10}x\sqrt{-g}\,\left[e^{-2 \phi} \left(R + 4 (\nabla \phi)^2\right) -\frac12 F_0^2\right]\,, \\
	\label{eq:SDBI}	S^\mathrm{DBI}&= -\sum \tau_i  \int_{\Sigma_i} d^9 x e^{-\phi}\sqrt{-h}\,,
\end{align}
where $\Sigma_i$ are the sources, and $h_{MN}$ is the metric induced on them by the bulk metric.

Our point of view in this subsection will be that the internal space
\begin{equation}
	M_5\times [0,z_0]\,,
\end{equation}
whose boundary consists of the loci where the two O8-planes sit. Thus we should be careful to include boundary terms. First of all, the variation of a bulk term in the action can yield a boundary term when we integrate by parts to extract its equations of motion. For example, the dilaton kinetic term in (\ref{eq:S0}), when varied with respect to $\phi$, upon integration by parts produces a bulk term that contributes to the equations of motion, but also a boundary term
\begin{equation}\label{eq:bd-dil-var}
	8\int_{\partial M_{10}} d^9 x \sqrt{-h}\, e^{-2 \phi} n^M \partial_M \phi\,,
\end{equation}
where $n^M$ is the normal vector to $\partial M_{10}$, normalized so that $n^2\equiv g^{NM}n_N n_M=1$. Second, as usual in general relativity, in presence of a boundary the action should also contain the Gibbons--Hawking--York (GHY) boundary term, even prior to variation. Varying the usual Einstein--Hilbert action produces a term containing the normal derivative of the metric variation $n^M \partial_M \delta g_{NP}$, which would require a restricted variational problem. The GHY term cancels this variation. In the Einstein frame it is equal to the integral of the boundary extrinsic curvature:
\begin{equation}\label{eq:GHYE}
	S^\mathrm{GHY,E} = \frac2{\kappa^2}\int_{\partial M_{10}} d^9 x \sqrt{-h_\mathrm{E}} \,\nabla^\mathrm{E}_M n^M_\mathrm{E}\,.
\end{equation}
In the string frame, a GHY-like term that achieves the same cancellation is 
\begin{equation}\label{eq:GHY}
	S^\mathrm{GHY} = \frac2{\kappa^2}\int_{\partial M_{10}}d^9 x \sqrt{-h} \,e^{-2 \phi}\nabla_M n^M\,.
\end{equation}
Notice that (\ref{eq:GHY}) does not immediately turn into (\ref{eq:GHYE}) upon the usual change of frame $g^\mathrm{E}_{MN} \equiv e^{-\phi/2} g_{MN}$. Rather, it does so when one combines it with a further boundary term produced by changing frame. Indeed
\begin{equation}
	\int_{M_{10}}d^{10}x \sqrt{-g_\mathrm{E}}\,\left(R_\mathrm{E}-\frac12 (\nabla \phi)^2 -\frac12 e^{\frac52 \phi} F_0^2\right) \equiv S^0_\mathrm{E} = S^0+ \frac9{2\kappa^2}\int_{\partial M_{10}} d^9 x\sqrt{-h} e^{-2 \phi} n^M \partial_M \phi\,,
\end{equation}
and 
\begin{align}
		S^\mathrm{GHY,E}&= \frac2{\kappa^2}\int_{\partial M_{10}}d^9x \frac{\sqrt{-h}}{\sqrt{-g}}e^{\frac14 \phi}\partial_M(e^{-\frac94 \phi}\sqrt{-g} g^{MN} n_N)= \nonumber\\
		&=\frac2{\kappa^2}\int_{\partial M_{10}} d^9x\sqrt{-h} \,e^{-2 \phi}\left[\frac1{\sqrt{-g}}\partial_M(\sqrt{-g} g^{MN} n_N)-\frac94 \partial_M \phi g^{MN} n_N\right]\\
		&=\frac2{\kappa^2}\int_{\partial M_{10}} d^9x\sqrt{-h}\,e^{-2 \phi}\nabla_M n^M-
		\frac9{2 \kappa^2}\int_{\partial M_{10}} d^9x\sqrt{-h}\,e^{-2 \phi} n^M \partial_M \phi\,. \nonumber
\end{align}
So $S^0+ S^\mathrm{GHY}= S^0_\mathrm{E}+S^\mathrm{GHY,E}$.

The total string-frame action is then
\begin{equation}
	S_{1/2}= S^0+ \frac12 S^\mathrm{DBI} + S^\mathrm{GHY} \, .
\end{equation}
(The subscript ${}_{1/2}$ reminds us that we are working with half of spacetime; the factor $1/2$ in front of the DBI action is present for the same reason.) Let us vary it with respect to the dilaton. The bulk variation produces  (\ref{eq:eom2}); but we are now interested in the boundary terms. The boundary contribution of $ \delta S_0$ is (\ref{eq:bd-dil-var});  $S^\mathrm{DBI}$ are already localized on the boundary, and also contribute. In particular notice that $\delta S^\mathrm{GHY}$ is non-zero because of the $e^{-2 \phi}$ factor, which is absent in the Einstein frame. Evaluating it explicitly in our metric ansatz (\ref{eq:metriceasy}) gives 
\begin{align}
	\delta_\phi S^\mathrm{GHY}&= -\frac4{\kappa^2}\int_{\partial M_{10}}d^9x \sqrt{-h}\delta \phi e^{-2 \phi} \, \nabla^M n_M 
	=-\frac4{\kappa^2} \mathrm{Vol}_{\mathrm{dS}_4} \mathrm{Vol}_{M_5} \delta \phi e^{-2 \phi+W} \partial_z(e^{-W+5 \lambda})\nonumber \\
	&= -\frac4{\kappa^2} \mathrm{Vol}_{\mathrm{dS}_4} \mathrm{Vol}_{M_5} \delta \phi e^{-2 \phi +5 \lambda} (5 \lambda-W)'|_{z\to z_0^+}\,
\end{align}	
taking into account that $n^2=1$. Combining all the contributions we obtain 
\begin{equation}\label{eq:bc-delta}
	\kappa^2 \delta_{\phi} S =  4 \mathrm{Vol}_{\mathrm{dS}_4} \mathrm{Vol}_{M_5} e^{- W
  + 5 \lambda - \phi} \delta \phi \left( e^{W - \phi} (2 \phi' + W' - 5
  \lambda') \mp \frac1{2\pi} \right)|_{z\to z_0^+}\,.
\end{equation}

At the O8$_+$, (\ref{eq:bc-delta}) is a linear combination of the (\ref{eq:bc-O+}). At the O8$_-$, from (\ref{eq:efi-O8-}) we see that the prefactor $e^{- W+ 5 \lambda - \phi}\sim\frac1{|z-z_0|}$:
\begin{equation}\label{eq:bc-delta-}
\begin{split}
	0&=\frac1{|z-z_0|}\delta \phi \left( e^{W - \phi} (2 \phi' + W' - 5
	  \lambda')+ \frac1{2\pi} \right)|_{z\to z_0^+}\\
	&=\frac1{|z-z_0|}\delta \phi \left( d + e|z-z_0|+\ldots\right) 
\end{split}\, \qquad (\text{O8}_-)\,.
\end{equation}
These $d$ and $e$ are related to our notation in (\ref{eq:deLR}). So now the interpretation of (\ref{eq:bc-delta-}) depends on the boundary condition one imposes on the fluctuation $\delta \phi$; or in other words, to what space $\delta\phi$ belongs. At first sight, a few possibilities might spring to mind: 
\begin{itemize}
	\item  $\delta \phi \in L^2(M_{10})$. Since $\sqrt{g}$ diverges as $\frac1{|z-z_0|^2}$ on the O8$_-$, this requires $\delta \phi \to 0$; if it goes like a power law, $\delta \phi \sim |z-z_0|^\alpha$, then $\alpha>1/2$. Then (\ref{eq:bc-delta-}) only requires setting the leading order $d=0$. So we are getting a linear combination of the permissive boundary conditions (\ref{eq:perm}).
	\item $\delta \phi$ smooth. In particular its limit for $z\to z_0$ can be a non-zero constant; this requires both the leading and subleading order, $d=e=0$. In this case we are getting a linear combination of the restrictive boundary conditions (\ref{eq:restr}). 
\end{itemize}
Intuitively, the more permissive we are with our variation space, the more restrictive the boundary conditions, because we are varying in more directions in field space. 

The two possibilities we have just seen are just some natural-sounding possibilities; others can be considered. The permissive boundary conditions are obtained more generally for any boundary condition that forces $\delta \phi\to 0$. 
 Making $\delta \phi \sim |z-z_0|^\alpha$, $\alpha>1$ would impose no boundary conditions at all; at the opposite extreme, leaving $\delta \phi$ completely unconstrained, free to diverge, would impose infinitely many boundary conditions. In section \ref{sub:kk} we will try to get a more physical picture of what might be reasonable conditions on $\delta \phi$. 

We are now going to look at the same issue from a slightly different viewpoint, using delta functions and distributions.


\subsection{Delta-Function Sources} 
\label{sub:delta}

We now consider the internal space as 
\begin{equation}
	M_5\times S^1\,,
\end{equation}
where $S^1= [0,2z_0]$ with the periodic identification $2z_0\cong 0$, and the functions are now required to be even under $z\to -z$.

From this point of view, there is no boundary, and no need to include boundary GHY terms; the action is now
\begin{equation}
	S=S^0 + S^\mathrm{DBI}
\end{equation}
with (\ref{eq:S0}), (\ref{eq:SDBI}). We can rewrite the localized DBI terms as a bulk term including a delta function, and then vary. This is how we found the equations of motion (\ref{eq:eom}), but the discussion so far (and in particular the difference between (\ref{eq:perm}) and (\ref{eq:restr})) made it clear that it is crucial to also understand how the variation multiplies it; so let us see this in more detail. We again focus on the dilaton equations of motion:
\begin{align}
		\kappa^2 \delta_\phi S&= -4 \int_{M_{10}}d^{10}x \sqrt{-g}\delta \phi e^{-2 \phi} (-2R-8 (\nabla e^{-\phi})^2) - \sum_i \tau_i \int_{\Sigma_i}d^9x\sqrt{-h}\delta \phi e^{- \phi} \nonumber\\
		&= -4\int dz e^{-W+5 \lambda-\phi} \delta \phi \left(e^{W-\phi}(W + 2 \phi - 5 \lambda)''+\ldots + \frac1\pi \sigma \right)\,. \label{eq:focus-dilaton}
\end{align}
The parenthesis is nothing but the equations of motion (\ref{eq:eom2}), and again the $\ldots$ are terms irrelevant for our discussion of what happens near the O8$_\pm$. Recalling (\ref{eq:efi-O8-}), we have obtained that (\ref{eq:eom2}) in fact arises as
\begin{equation}\label{eq:bc-delta-2}
	\frac1{|z-z_0|} \delta \phi \left(e^{W-\phi}(W + 2 \phi - 5 \lambda)''+\ldots + \frac1\pi \delta \right)=0  \qquad (\text{O8}_-)\,.
\end{equation}
This is the delta-function counterpart of (\ref{eq:bc-delta-}). Once again we see that the conditions we impose on $\delta \phi$ play a crucial role. Mirroring our discussion in the previous subsection:
\begin{itemize}
	\item $\delta \phi\in L^2(M_{10})$. Then $\delta \phi\to 0$; in this case the second derivatives produces a delta term to match the explicit $\delta$ in the parenthesis; the subleading terms in the $|z-z_0|$ expansion evaluate to zero when multiplied by the prefactor $\frac{ \delta \phi}{|z-z_0|}$.  Thus in this case we only have a condition on the leading behavior. 
	\item $\delta \phi$ smooth. This in particular allows $\delta \phi$ to go to a constant. Then we have a term $\frac1{|z-z_0|} \delta(z-z_0)$, as anticipated in (\ref{eq:1/z-delta}). This is of unclear interpretation, but if it can be given a meaning, it is likely to require two conditions on $e^{W-\phi}(W + 2 \phi - 5 \lambda)''$ and thus on the functions: not just a condition on their leading behavior, but on their subleading behavior as well.
\end{itemize}
So we recover the issue we saw in the previous subsection, although with the disadvantage of having products of distributions. This is expected as General Relativity is a non-linear theory, and only in very particular cases the field equations become linear allowing for a rigorous treatment of the singularities within the framework of usual linear distribution theory.
This is the case for the flat-space solutions described in section \ref{sec:op}, where the field equations reduce to a Gauss-like equation for the harmonic function $H$ of the form $\Delta H = \tau \delta$.
In more general cases, like the present one, the non-linearity of the field equations introduces ill-defined products of distributions.\footnote{Such products of distributions could perhaps be defined in a more general mathematical framework.
One such approach is based on Colombeau algebras, which include distributions as a linear subspace and smooth functions as a subalgebra.
For a review of applications of these methods in General Relativity see for example \cite{Steinbauer:2006qi} and references therein.
It would be interesting to apply these methods to the present problem.}

Given the problems we have just seen with interpreting the $\delta$ terms, we might wonder whether perhaps we are working in the wrong set of variables. Perhaps the issues are created by the fact that the $f_i$ are diverging, and it might be wiser to switch to variables that remain finite. 

For example we can define 
\begin{equation}
	H_i = e^{-4 f_i}\,;
\end{equation}
the metric then reads
\begin{equation}
  d  s^2_{10} = H_1^{- 1 / 2} d  s^2_{d  S_4} + H_1^{1
  / 2} (d  z^2 + H_3^{- 1} d  s^2_{M_5})\,.
\end{equation}
The equations of motion are now 
\begin{subequations}\label{eq:eom-H}
\begin{equation}
	  \Lambda  =  \frac{F_0^2  \sqrt{\frac{H_1}{H_2}}}{8 H_1 H_2^2} - \frac{5
	  H_3 \kappa}{4 H_1} - \frac{(H_1')^2}{8 H_1^3} + \frac{25 (H_2')^2}{16 H_1
	  H_2^2} + \frac{5 (H_3')^2}{4 H_1 H_3^2} + \frac{5 H_1' H_2'}{16 H_1^2 H_2} -
	  \frac{25 H_2' H_3'}{8 H_1 H_2 H_3}
\end{equation}
\begin{equation}
	 \pm \frac4\pi \sigma  =  - \frac{H_2^{5 / 4} H_1''}{H_1^{5 / 4}} -
	  \frac{F_0^2  \sqrt[4]{H_1}}{H_2^{5 / 4}} - 4 H_1^{3 / 4} H_2^{5 / 4} \Lambda
	  + \frac{H_2^{5 / 4} (H_1')^2}{H_1^{9 / 4}} - \frac{5 \sqrt[4]{H_2} H_1'
	  H_2'}{2 H_1^{5 / 4}} + \frac{5 H_2^{5 / 4} H_1' H_3'}{2 H_1^{5 / 4} H_3}
\end{equation}
\begin{equation}
	  \pm \frac4\pi \sigma  =  - \frac{H_2^{5 / 4} H_3''}{\sqrt[4]{H_1} H_3} -
	  \frac{F_0^2  \sqrt[4]{H_1}}{H_2^{5 / 4}} - \frac{2 H_2^{5 / 4} H_3
	  \kappa}{\sqrt[4]{H_1}} - 2 H_1^{3 / 4} H_2^{5 / 4} \Lambda + \frac{7 H_2^{5
	  / 4} (H_3')^2}{2 \sqrt[4]{H_1} H_3^2} - \frac{5 \sqrt[4]{H_2} H_2' H_3'}{2
	  \sqrt[4]{H_1} H_3}
\end{equation}
\begin{align}
	  \pm \frac4\pi \sigma & =  - \frac{\sqrt[4]{H_2} H_2''}{\sqrt[4]{H_1}} -
	  \frac{4 F_0^2  \sqrt[4]{H_1}}{5 H_2^{5 / 4}} - \frac{2 H_2^{5 / 4} H_3
	  \kappa}{\sqrt[4]{H_1}} - \frac{8}{5} H_1^{3 / 4} H_2^{5 / 4} \Lambda -
	  \frac{H_2^{5 / 4} (H_1')^2}{5 H_1^{9 / 4}} + \frac{2 H_2^{5 / 4}
	  (H_3')^2}{\sqrt[4]{H_1} H_3^2}  \nonumber\\
	  &   + \frac{\sqrt[4]{H_2} H_1' H_2'}{2 H_1^{5 / 4}} - \frac{5
	  \sqrt[4]{H_2} H_2' H_3'}{2 \sqrt[4]{H_1} H_3} +
	  \frac{(H_2')^2}{\sqrt[4]{H_1} H_2^{3 / 4}}\,. 
\end{align}	
\end{subequations}
Near the O8$_-$, (\ref{eq:efi-O8-}) tells us that $H_i\sim c_i^{-4} |z-z_0|$. Looking at (\ref{eq:eom-H}), we see that the coefficients of $H_i''$ are all constant; taking this into account, near the O8$_-$ now the equations are of the form $\del_z^2 H_i= \delta$ rather than the more confusing (\ref{eq:eom-dots}). But notice that this agreement only required the leading behavior in $z-z_0$; so from this point of view the permissive boundary conditions seem to be enough to reproduce the delta's in the equations of motion.


\subsection{Finite Masses} 
\label{sub:kk}

We have seen in detail how the problem of the boundary conditions can be traced back to the boundary conditions for the field fluctuations. Let us now look at a possible strategy one might try to use to decide the correct conditions for the field fluctuations. 

Namely, one might try to perform a KK reduction on a solution. We work in the point of view where the internal space is $M_5\times S^1$, as in section \ref{sub:delta}. Unfortunately this is a very convoluted computation, but we can try to look at a small block in the mass matrix; once again we will consider the dilaton fluctuations $\delta \phi$. The second variation of the action reads
\begin{equation}\label{eq:d2S}
	\delta^2_\phi S = \frac8{\kappa^2}\int_{M_{10}}d^{10}x \sqrt{-g} e^{-2 \phi} \nabla_m \delta \phi \nabla^m \delta \phi +\sum_i \tau_i \int_{\Sigma_i} d^9 x \sqrt{-h}e^{-\phi} \delta \phi^2\,.
\end{equation}
We now expand the dilaton perturbation on not-yet-specified basis of functions in the internal space: 
\begin{equation}
	\delta \phi = \sum_k \varphi_k (x) f_k(y)\,,
\end{equation}
where $x$ and $y$ denote external and internal coordinates respectively. Plugging this into (\ref{eq:d2S}) we get
\begin{align}\label{eq:d2S-exp}
	&\delta^2_\phi S =  \int_{M_4} \sqrt{-g_4}d^4 x \Big[ g_4^{\mu \nu} \partial_\mu \varphi_i \partial_\nu \varphi_j \frac8{\kappa^2}\int_{M_5} d^5 y \int dz  \sqrt{g_5} e^{-4W + 5 \lambda - 2 \phi} f_i f_j  \\
	&+ \varphi_i \varphi_j  \int_{M_5} d^5 y \sqrt{g_5}\Big(e^{-W + 5 \lambda - \phi} f_i f_j + \frac8{\kappa^2}\int dz\Big(  
		e^{5 \lambda - 2 \phi} f_i' f_j' + e^{3 \lambda - 2 \phi} g_5^{ab} \partial_a f_i \partial_b f_j\Big)\Big)
		\Big]\,. \nonumber
\end{align}
Now the $\varphi_i$ are interpreted as four-dimensional scalar fields. The first line in (\ref{eq:d2S-exp}) gives their kinetic terms, while the second line gives their mass matrix. We see that some of the terms might diverge. Let us consider for example the boundary condition where $f$ is taken to be smooth, which as we saw earlier leads to the restrictive boundary conditions. In this case, all the terms containing an integral in $dz$ converge; but the term $e^{-W + 5 \lambda - \phi} f_i f_j$, which comes from the localized term in (\ref{eq:d2S}), goes like $\frac1{|z-z_0|} f_i f_j$ and hence diverges for $f_i$ smooth. So the restrictive boundary conditions in this case lead to a diverging mass matrix, which is presumably unphysical. 

On the other hand, if we take $f_i \sim |z-z_0|^\alpha$ with $\alpha > 1/2$, which according to (\ref{eq:bc-delta-}) leads to the permissive boundary conditions, then this block in the mass matrix is finite. 

We hasten to add, however, that this is only a very small piece of the KK reduction. In order to compute the mass matrix, one actually needs to first make sure all the fields with different spins in four dimensions to decouple, and this might change the block of the mass matrix we have computed above. It is possible to imagine that other naively divergent terms appear, and that they combine with the one we have discussed here to give a finite mass. We even saw a possible indication of this in the analogue of (\ref{eq:d2S-exp}) in the Einstein frame. Even if this happens, however, such a cancellation of infinities seem to depend on the ``scheme'' one chooses to regularize the infinities, similar to our discussion of the on-shell action in section \ref{sub:action}. 

To summarize: depending on the boundary conditions on the fluctuations, one will obtain a larger or smaller set of fields in four dimensions. The restrictive boundary conditions on the fields correspond to a laxer condition on the fluctuations, which would result in more four-dimensional scalar fields $\varphi_i$. The 4d equations of motion for the additional fields obtained in this way  would presumably not be obeyed, thus invalidating our solutions from another point of view. However, we found in this subsection that these putative additional $\varphi_i$ seem in fact to have an infinite mass matrix.


\subsection{On-Shell Action} 
\label{sub:action}

In the previous subsection we examined our solutions from the point of view of the four-dimensional effective action. We focussed on  small fluctuations, interpreting them as scalars in four dimensions. In this section, we will focus instead on the parameter $c$ in (\ref{eq:resc}). 

If stringy corrections are not considered, $c$ appears to be a modulus of our solutions. As we discussed earlier, (\ref{eq:resc}) is only a symmetry of the supergravity equations of motion, and not of those of full string theory; so $c$ will be fixed in the full theory. In other words, there will be an effective potential $V(c)$; the question of existence of our solutions in full string theory is the same as the existence of an extremum of this potential. $V(c)$ will have contributions from all stringy corrections, and for this reason it is difficult to compute. In a way the complaint in \cite{cribiori-junghans} amounts to saying that supergravity also gives a contribution to $V(c)$. As we discussed, however, in the strongly-coupled region supergravity is the least important term in the equations of motion, and so its contribution will be completely swamped by more important ones. 

However, the spirit of this section has been to examine the formal problem of the existence of the solutions in supergravity. In that spirit, let us try to see how one should interpret $c$. If it can be interpreted as a scalar field in four dimensions, then it would enter the four-dimensional action:
\begin{equation}\label{eq:S4d}
	S_{4d} = \int \sqrt{-g_4}(R_4 - V(c))\,.
\end{equation}
However, we quickly see a puzzle. The rescaling (\ref{eq:resc}) acts on the four-dimensional metric as $g_{\mu \nu}\to e^{2c}g_{\mu \nu}$. The Ricci scalar $R_4\to e^{-2c}R_4$. For a shift in $c$ to be a symmetry, we would need the potential to rescale in the same way, $V(c)= V_0 e^{-2c}$. This does not seem to be compatible with a potential that has a vacuum, unless the constant $V_0=0$, i.e.~the vacuum is Minkowski. 

The resolution of this puzzle is clear given our discussion in section \ref{sub:kk}: \emph{$c$ cannot be viewed as a field}. Indeed, the infinitesimal counterpart of (\ref{eq:resc}) corresponds to fluctuations 
\begin{equation}
	\delta g_{\mu \nu}= 2  g_{\mu \nu} \delta c \, ,\qquad \delta g_{mn}= 2 g_{mn} \delta c \, ,\qquad\delta \phi = -  \delta c\,.
\end{equation}
Looking at our (\ref{eq:bc-delta-}) and our discussion there, we see that this would require a boundary condition where the two leading coefficients $d$ and $e$ should be set to zero; this is the restrictive condition (\ref{eq:restr}). Similar conclusions can be reached with (\ref{eq:bc-delta-2}). So when $c$ can be considered as a field, our solutions don't exist; this is consistent with (\ref{eq:S4d}). 

On the other hand, with the boundary conditions (\ref{eq:perm}) we have imposed, $\delta c$ is not part of the space of allowed field variations. 

What if we try to evaluate (\ref{eq:S4d}) directly?\footnote{We thank J.~Maldacena for discussions on this point.}  After all, we should be able to compute its value by integrating on the solution the ten-dimensional $S$ over $M_6$, for any value of $c$. (Equivalently, we can integrate $S$ over $M_{10}$; this would diverge because of the volume of de Sitter space, but we can take care of this by analytically continuing to Euclidean signature.) 

Perhaps unsurprisingly at this point, the answer is that the Lagrangian density is divergent at the O8$_-$, and so this on-shell action is ill-defined. The computation is similar to (\ref{eq:focus-dilaton}). The bulk action (\ref{eq:S0}) diverges on shell as $\int dz \frac1{|z-z_0|^2}$:
\begin{equation}\label{eq:S0-onshell}
\begin{split}
		\kappa^2 S^0\propto &\int d z \Big[2e^{5 \lambda -2 \phi} \Big((5 \lambda-W)'' \\ 
		&+ 4 (W')^2 -5 W' \lambda' + 15 (\lambda')^2 -2 (\phi')^2 \Big) - \frac12 e^{5 \lambda -2 W} F_0^2 \Big]+ \ldots\,.     
\end{split}
\end{equation}
The O8$_-$ term is even more puzzling:
\begin{equation}\label{eq:SDBI-onshell}
	 S^\mathrm{DBI} \propto e^{-W+5 \lambda - \phi}|_{z=z_0} =\int dz e^{-W+5 \lambda - \phi} \delta(z-z_0) \,,
\end{equation}
so it again would involve evaluating $\frac1{|z-z_0|} \delta(z-z_0)$, an expression which has plagued our discussion since (\ref{eq:1/z-delta}). 

It is possible to find regularization schemes that make the divergences in (\ref{eq:S0-onshell}) and (\ref{eq:SDBI-onshell}) cancel each other, and leave a finite answer. For example, we can decide to introduce a length cutoff: we can evaluate the integral in (\ref{eq:S0-onshell}) only up to $z_0-\epsilon$, and interpret (\ref{eq:SDBI-onshell}) as $e^{-W+5 \lambda - \phi}|_{z=z_0-\epsilon}$. The divergences in $S^0 + S^\mathrm{DBI}$ now do cancel, and leave a finite result, which one might try to interpret as (\ref{eq:S4d}). However, this regularization scheme is highly arbitrary. It is equivalent to regularizing $\delta(z-z_0)$ as $\frac12 \left(\delta(z-z_0+ \epsilon) + \delta(z-z_0 - \epsilon) \right)$. More conventional regularizations for $\delta(z-z_0)$, where for example one replaces it with a Gaussian of width $\epsilon$, would give different results, or fail to cancel the divergence altogether.

So a direct attempt at computing (\ref{eq:S4d}) by regularizing $S=S^0+S^\mathrm{DBI}$ seems to fail, and to give a highly ambiguous result. Once more, supergravity fails to decide by itself how it should be defined at strong coupling.


\subsection{Summary} 
\label{sub:summary}

At the risk of repeating ourselves, we summarize here the results of our discussion. 

The solutions in \cite{cordova-deluca-t-ds4} were obtained by using the supergravity equations of motion; they displayed a singular behavior that was the same as an O8$_-$ in flat space at leading order in the distance $|z-z_0|$ from it. Since the supergravity approximation breaks down near this singularity, however, we could not establish in \cite{cordova-deluca-t-ds4} whether the solution survives in full string theory. 

On the other hand, \cite{cribiori-junghans} used the supergravity equations of motion to say that the solutions don't make sense even in supergravity. We think this has no bearing on the issue of whether they exist in string theory, which is the physically meaningful question. 

However, in this section we have tried to assess this claim. We have argued that:
\begin{itemize}
	\item There are several versions of the O8$_-$ boundary conditions. In particular, in \cite{cordova-deluca-t-ds4} we imposed a permissive version, while \cite{cribiori-junghans} in their criticism implicitly used a more restrictive version. 
	\item The choice between permissive and restrictive version is in turn related to the choice of what field fluctuations we allow near the O8$_-$. A laxer condition on the field fluctuations leads to restrictive boundary conditions, and vice versa. 
	\item From a KK point of view, this can also be interpreted as the inclusion of more or fewer scalars in four dimensions. 
	\item A modulus of the solutions in \cite{cordova-deluca-t-ds4}, which is expected to be lifted by string corrections, cannot be interpreted as a field in four dimensions because it would correspond to fluctuations which do not vanish on top of the O8$_-$ and thus are not allowed by the permissive boundary conditions. 
\end{itemize}

So one cannot really decide in supergravity alone whether the solutions in \cite{cordova-deluca-t-ds4} make sense. The issue depends on an ambiguity that manifests itself at various levels in the theory, and which has to do with how to interpret the theory near a strongly coupled O8-plane. Ultimately it just signals that supergravity is not well-defined by itself at strong coupling, and needs a UV-completion. This confirms that we need string theory to decide the fate of the solutions in \cite{cordova-deluca-t-ds4}. We do note again, however, that supersymmetric solutions exist in the literature which satisfy the permissive boundary conditions used there; they are reviewed in appendix \ref{bdsusy}. 



\section{Conclusions} 
\label{sec:conclusions}

In this paper, we have obtained new dS$_4$ solutions with an O8$_+$ and an O6$_-$, and we have reexamined the validity of our older ones in \cite{cordova-deluca-t-ds4}. For both solutions, the presence of O-planes is inferred by comparison with their flat-space behavior. Since the latter have strong curvature and coupling, stringy corrections come into play, and it is impossible to decide with supergravity alone whether the solutions are valid. It is important to stress that this will be so for \emph{any} solution with O-planes.

It would be important, then, to develop techniques to decide whether a solution with O-planes will survive in full string theory. In other words, it would be important to understand what conditions one needs to impose near the O-plane singularities.\footnote{The nature of these singularities is also important for the KK spectrum, as emphasized in section \ref{sub:kk} and for example in \cite{passias-richmond,andriot-tsimpis-gw}.}
For example, for the O8$_+$--O6$_-$ solutions of section \ref{sec:o8o6}, it is possible --- perhaps even likely --- that there are some extra physical requirements one needs to impose. We tried to impose a condition based on a formal analytic continuation, but this was only based on analogies and not well justified. For the O8$_+$--O8$_-$ solutions in section \ref{sec:o8o8}, we have examined two possibilities (dubbed ``restrictive'' and ``permissive''), with neither emerging as a clear winner. We clearly need alternative procedures that are better justified physically.

A first possibility is to probe the singularity with D-branes. 
There are various possible ways to do so. For example, for the massless O6 singularity, a probe D2 gives useful information. The idea here is that the backreaction of a D-brane or O-plane is generated by integrating out open string modes. So by computing the quantum effective theory on a probe D2 we are really computing what the backreaction should be in full string theory. 
The low-energy effective action of a D2 in an O6$_-$ background is $\mathrm{SU}(2)$ with ${\mathcal N}=4$ supersymmetry (in $d=3$). 
Perturbatively, the metric on its Coulomb moduli space would have a singularity and cease to make sense near the origin; fortunately, instanton corrections modify it and turn it \cite{seiberg-witten-3d,dorey-khoze-mattis-tong-vandoren} into a smooth hyper-K\"ahler space called ``Atiyah--Hitchin'' manifold AH$_4$ \cite{atiyah-hitchin}. In string theory, this is interpreted as the statement that the O6 singularity is resolved in M-theory to $\mathbb{R}^7\times \mathrm{AH}_4$ \cite{seiberg-witten-3d,hanany-pioline}.

Something similar might be attempted for the O6$_-$ in presence of Romans mass $F_0$.  One step in this direction was done in \cite{saracco-t-torroba}. Because of the coupling $F_0\int \mathrm{CS}(a)$ on a D2 (where $a$ is the worldvolume gauge potential), one expects the effective D2 action to now include a Chern--Simons (CS) term; \cite{saracco-t-torroba} then computed the Coulomb branch metric on an ${\mathcal N}=2$ $\mathrm{SU}(2)$ CS theory, finding a behavior in qualitative agreement with the smooth behavior found in \cite{saracco-t} for a certain class of O6 solutions with $F_0$. This analysis is incomplete because the Lagrangian in \cite{saracco-t-torroba} was not fully justified in string theory; adding a CS term to a supersymmetric theory in three dimensions can be done in several different ways, and each of them can potentially lead to different behaviors of the Coulomb branch metric, corresponding probably to the geometry parallel to the O6. (Indeed one expects so, since the O6 behavior in the supersymmetric solutions of section \ref{sub:ads7} have a quite different behavior from those in \cite{saracco-t}.) Performing this computation carefully might reveal what one should really expect from the quantum O6 singularity. 

Another possible logic might be the one in \cite{michel-mintun-polchinski-puhm-saad}. This regards the backreaction of an object on itself, viewed as an effective field theory. This might not be appropriate for O-planes, which are non-dynamical, although we remark that one of the examples in \cite{michel-mintun-polchinski-puhm-saad} is the back-reaction of a defect coupled to a bulk scalar and with no localized degrees of freedom. 

One might try to compute the tension associated to our singularities without going on top of them. After all, in general relativity we usually don't compute the mass of a gravitational source by checking the delta in the equations of motion (although for some early attempts in this direction see \cite{Balasin:1993fn, Balasin:1993kf}). Rather, we compute the gravitational field far away from it. Various formalizations of this procedure exists, including the Komar and ADM mass. Unfortunately for us these are not very relevant, since we cannot go far away from the sources, given that the internal space is compact. A more promising alternative is the covariant phase space formalism (for a review see \cite{compere-fiorucci}), which in principle reduces the computation to a Gauss-like integral. It would be interesting to develop this further.
 
Finally, an indirect way of testing singularities is to use holography for AdS solutions that incldue them. For example, the solutions reviewed in appendix \ref{bdsusy} have O8$_-$ singularities with the permissive boundary conditions of \cite{cordova-deluca-t-ds4} and are supersymmetric; finding their CFT duals would presumably settle the issue. Regarding O6$_-$ singularities, there are AdS$_7$ solutions which include them \cite{afrt} and which have been tested holographically \cite{apruzzi-fazzi}, which we have in fact used as a starting point for our new solutions in section \ref{sec:o8o6}. For these one could perhaps obtain stronger checks by going to subleading orders in $N$.


\section*{Acknowledgements}

We thank N.~Cribiori, K.~Eckerle, D.~Junghans, G.~Lo Monaco, J.~Maldacena and A.~Passias for discussions. GBDL and AT are supported in part by INFN. AT is also supported in part by the MIUR-PRIN contract 2017CC72MK003.

\appendix

\section{AdS$_4$ $\mathcal{N} = 2$ Solutions with Permissive O8$_{-}$s} 
\label{bdsusy}

In this section we review some analytic AdS$_4$ solutions \cite{passias-prins-t} of massive IIA with $\mathcal{N} = 2$ supersymmetry. They were found as part of a larger class which also allows fully regular solutions; the ones which are relevant here can be found in \cite[Sec.~4.1.4, 4.2.3, 4.2.4]{passias-prins-t}. The solutions have an O8$_{-}$ with the permissive boundary conditions of \cite{cordova-deluca-t-ds4}, and no other source.

The ten-dimensional metric is
\begin{equation}
  d  s^2_{10} = e^{2 W} d  s^2_{\text{AdS}_4} + e^{- 2 W}
  (e^{2 Q} d  x^2 + e^{2 \lambda_1} D \psi^2 + d s^2_4 ) \;,
\end{equation}
where all the functions only depend on the coordinate $x$. The coordinate $\psi$ parametrizes an $S^1$ fibered over the internal
space with a connection $\rho$, and $D \psi \equiv d \psi + \rho$. The warping functions are determined in terms of single
polynomial $q (x)$ as
\begin{equation}
		e^{2 W} = L^2 x \sqrt{1 - \frac{4q}{xq'}}
		\,,\qquad 
		e^{2 Q} = 
	  L^4\left( 1-\frac{x q'}{4q}\right) \, ,\qquad
		e^{2 \lambda_1}=- L^4  \frac{x q}{q'} \, .
\end{equation}
The only fluxes present are $F_0$ and 
\begin{equation}
	F_4= -F_0 L^4 e^{-2A}(A'x-1) p dx \wedge D \psi \wedge j +\frac12 F_0 (4A'x+1)p^2 j^2\,
\end{equation}     
where $p=\frac{\kappa}{6x}L^4 e^{-2A}(1-x^3)$. The four-dimensional metric $ds^2_4$ can be either proportional to a single K\"ahler--Einstein space  or to a product of two Riemann surfaces $\Sigma_1$ and $\Sigma_2$. In both cases, $j$ is the K\"ahler form of the metric $ds^2_4$.  In the K\"ahler--Einstein case, a possible potential $C_3$ such that $F_4 = d C_3$ is given by 
\begin{equation}
    C_3 = -\kappa F_0 L^2 \frac x2 \frac{x  (q')^2 + 4q (2q'-xq'')}{(x q''-3 q')^2} D \psi \wedge j\,. 
\end{equation}

In the first case, 
\begin{equation}
  ds^2_4 = \kappa L^4 x\frac{4q-x q'}{xq''-3q'} ds^2_{\mathrm{KE}_4} \,,
  \qquad e^{4 \phi} = \frac{2^6\cdot 3^4}{F_0^4 L^4} \frac{x^5}{q'} \frac{\left(x q' - 4 q\right)^3}{\left(x q''-3q'\right)^4}\,,
\end{equation}
where $\kappa$ is the sign of the curvature of the K\"ahler--Einstein space and the polynomial $q$ is given by 	
\begin{equation}
	q= x^6 + \frac \sigma2 x^4 + 4 x^3 -\frac12\,.
\end{equation}
    In the second case
\begin{equation}
	ds^2_4 =  \frac{L^4}{12} \left( \frac{x q' - 4 q}{x  (1-x^3)}  \kappa_1 ds^2_{\Sigma_1}
    -\frac{x q' - 4 q}{x  (t - x^3)} \kappa_2 ds^2_{\Sigma_2}\right)\,,\quad
    e^{4 \phi} = \frac{(F_0 L)^{-4}}{ 4 x^3 q'}\frac{\left(xq'-4q\right)^3}{\left(x^6-(1+t)x^3+t\right)^2}
\end{equation}
where $\kappa_1$ and $\kappa_2$ are the signs of the curvatures of $\Sigma_1$ and $\Sigma_2$ respectively
and the polynomial now depends on two parameters $\sigma$, $t$:
\begin{equation}
  	q = x^6 + \frac{\sigma}{2} x^4 + 2 (1 + t) x^3 - \frac{t}{2} \label{eq:polyq-pro} \;.
\end{equation}

Depending on the zeros of $q$ and
$q'$ different endpoints can be obtained. In particular, we are interested in
solutions with a regular point and an O8-plane. In order to have a regular
point at $x = x_0$ we need $q (x_0)$ to vanish linearly, such that $q' (x_0)
\neq 0$. Indeed, near such a point the metric behaves as
\begin{equation}
  d  s^2_{10} \sim d  s^2_{\text{AdS}_4} + \frac{1}{| x - x_0
  |} d  x^2 + | x - x_0 | D \psi^2 + ds^2_4  \,,
\end{equation}
which with the simple change of coordinates $r^2 \equiv | x - x_0 |$ is
equivalent to
\begin{equation}
  d  s^2_{10} \sim d  s^2_{\text{AdS}_4} + d  r^2 +
  r^2 D \psi^2 + ds^2_4\,.
\end{equation}
From the above expression we see that the $S^1$ parametrized by $\psi$ shrinks
regularly provided $\psi$ has the correct periodicity.
 
\begin{figure}[ht!]
\begin{center}
  \resizebox{281pt}{180pt}{\includegraphics{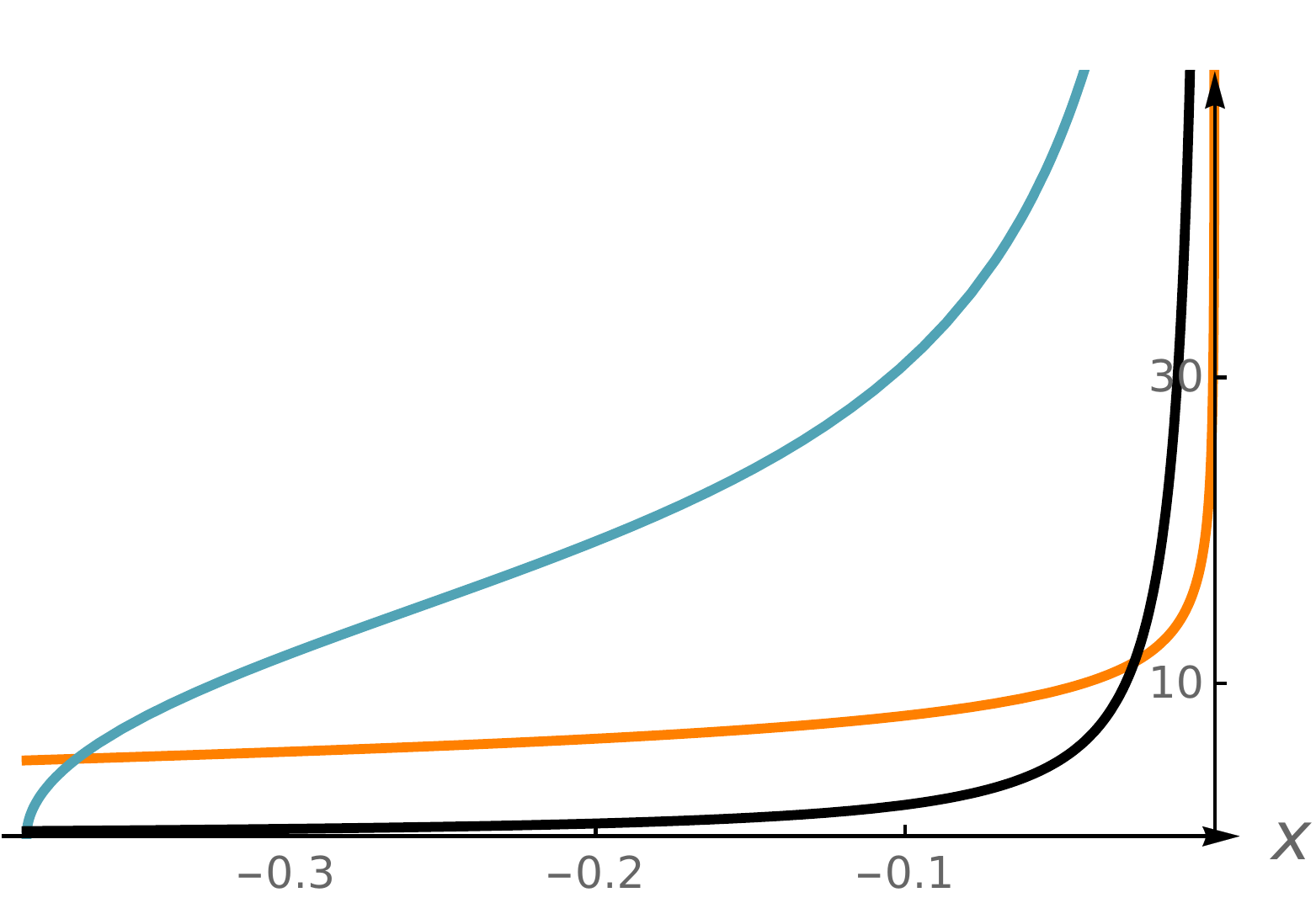}}
  \caption{A solution with $t = - \frac{1}{4}$, $\sigma = - 4$, $F_0 = -
  \frac{4}{2 \pi}$ , and $L = 8$. The plotted functions are $e^W$(orange),
  $e^{\phi}$(black) and $e^{\lambda_1}$(turquoise). On the left ($x_0 = -
  0.38$) the $S^1$ shrinks regularly, on the right ($x = 0$) a
  diverging-dilaton O8 sits.\label{fig:N2sol}}
\end{center}
\end{figure}

There are several cases where these boundary conditions on $q$ can be met, and we obtain solutions with a single O8$_-$ source:
\begin{itemize}
	\item In the case with a single K\"ahler--Einstein space, when the parameter $\sigma \ge -9$, the solution has an O8$_-$ at $x=0$ and caps off regularly at $x=x_0>0$. 
	\item For $-1<t<0$, $\sigma<3(-t)^{-1/3}(1+2t)$, $\kappa_1=1$, $\kappa_2=-1$, the solution caps off regularly at $x=x_0<0$ and has an O8$_-$ at $x=0$. An example of solution in this class is given in Figure \ref{fig:N2sol}. 
	\item For $t<-1$, $\sigma < -3 (2+t)$, $\kappa_1=1$, $\kappa_2=-1$, the solution has an O8$_-$ at $x=0$ and caps off regularly at $x=x_0>0$. 
	\item For $t>0$, $\sigma < -3 t^{-1/3}(1+2t)$, $\kappa_a=1$, the solution has an O8$_-$ at $x=0$ and caps off regularly at $x=x_0>0$.
	\item Finally the case $\kappa_2=0$ has to be treated differently; the equations for $ds^2_4$, $e^\phi$ and $q$ change. Here one can obtain solutions with a single O8$_-$ for $\kappa_1>0$.  
\end{itemize}

On these solutions we can compute the quantity $e^{5 W - \phi}$ analytically and we obtain
\begin{equation}
  e^{5 W - \phi} =\left\{
    \begin{array}{l}
      F_0 L^6 \frac{\left(x^3-1\right)^2}{3x^3+\sigma x +6}\\
      \\
      F_0 L^6 \frac{(t-x^3)(x^3-1)}{3x^3 + \sigma x+3(t+1)}
  \end{array}\right.
\end{equation}
so that in both cases
\begin{equation}
  \lim_{x \rightarrow 0} \partial_x(5W-\phi) \propto \sigma \neq 0\;.
\end{equation}
We see that generically this is non-zero, and hence satisfies the permissive boundary conditions (\ref{eq:perm}), but not the restrictive ones (\ref{eq:restr}).\footnote{A similar feature is present also for the O8$_-$ singularity in the AdS$_3$ solutions of \cite[Sec.~6.3]{dibitetto-lomonaco-passias-petri-t}, although those have three-dimensional ${\cal N}=1$ supersymmetry and hence only two supercharges.}

\bibliography{at}
\bibliographystyle{at}

\end{document}